\documentclass[letterpaper, 12pt]{article}

\usepackage{amsmath,amssymb,mathrsfs}
\usepackage{graphicx}
\usepackage{color}
\usepackage{ushort}
\usepackage{url}
\usepackage{lscape}
\usepackage{subcaption}
\usepackage{pdflscape}
\usepackage{bbm}
\usepackage{epstopdf}
\usepackage{amsthm}
\usepackage[onehalfspacing]{setspace}
\usepackage{natbib}
\usepackage{enumerate,enumitem}
\usepackage{libertine}
\usepackage[T2A, T1]{fontenc}
\usepackage[utf8]{inputenc}
\usepackage[english]{babel}
\usepackage{soul}
\usepackage{tcolorbox}
\usepackage{xcolor,colortbl}
\usepackage{threeparttable,multirow}
\usepackage{xfrac}
\usepackage{threeparttable}
\usepackage{rotating}
\usepackage{tabularx,ragged2e,booktabs}
\usepackage{subfiles}
\usepackage{amsmath}
\usepackage{amssymb}
\usepackage[margin=1in]{geometry}
\usepackage{bm}
\usepackage[title]{appendix}
\usepackage{setspace}

\definecolor{red}{rgb}{0.9,0.1,0.4}

\DeclareMathOperator{\supp}{supp}

\DeclareMathOperator{\argmax}{argmax}

\newtheorem{proposition}{Proposition}
\newtheorem{corollary}{Corollary}
\newtheorem{lemma}{Lemma}
\newtheorem{theorem}{Theorem}
\newtheorem{assumption}{Assumption}

\newtheorem{definition}{Definition}

\newtheorem{claim}{Claim}

\definecolor{Blue}{RGB}{0,32,216}
\usepackage[colorlinks=true,allcolors=Blue]{hyperref}
\usepackage{xcolor}
\usepackage{rotating}
\renewcommand{\cite}{\citet*}
\title{Contracting against a Non-contractible Outsider\thanks{I am grateful to Tan Gan, Marina Halac, and Elliot Lipnowski for their helpful discussions and comments.}}
\author{Hongcheng Li\thanks{
        Hongcheng Li: Department of Economics, Yale University (e-mail: \href{mailto:hongcheng.li@yale.edu}{hongcheng.li@yale.edu}).}}
                  
\date{\today}

\begin{document}
\onehalfspacing
\maketitle

\begin{abstract}
    \noindent I study a general framework of contracting with externalities involving a non-contractible outsider. \emph{Strategic symmetry} between the insider agent and the outsider is the source of multiple equilibria, even when their actions are strategic substitutes. To tackle strategic uncertainty, the principal guarantees a unique equilibrium outcome. A novel duality approach reformulates her problem as the choice of a series of agent expectations about outsider behavior, one for each agent action. The key constraint comes from eliminating undesired equilibria: she cannot convince the agent to expect higher incentives than those generated by the outsider's best responses to the actions she has already conquered. To reduce strategic rents, the principal optimally induces attenuated agent incentives. In supermodular or submodular environments, her coordination power and commitment power are substitutes in determining the contracting outcome, and perfect substitutes under further conditions. On contractual privacy, the principal can strictly prefer private contracting over public contracting, in sharp contrast to the benchmark that ignores robustness. Applications include regulating international competition, platform design, and labor union contracting.
\end{abstract}



\newpage
\tableofcontents
\newpage

\section{Introduction}\label{Introduction}

Contracts often lack contingencies on the decisions of third parties. This gap becomes a significant problem when the contracting parties and these outsiders face multilateral externalities. For example, a government offering subsidies through industrial policies to protect domestic firms\footnote{\cite{ejmr2024} and a \href{https://www.mckinsey.com/capabilities/geopolitics/our-insights/from-protection-to-promotion-the-new-age-of-industrial-policy}{2025 McKinsey article} document the widespread post-pandemic return of industrial policies, with subsidies as the most used instrument and advanced economies as the primary drivers.} may find the use of tariffs to tax away international competition largely constrained, because tariffs have limited flexibility and are hard to put into effect\footnote{For example, the \href{https://eur-lex.europa.eu/eli/treaty/tfeu_2016/art_30/oj/eng}{Treaty on the Functioning of the European Union} (Articles 28 and 30) forbids member states to impose internal tariffs on each other. Also, according to a recent report by the WTO (\href{https://www.wto.org/english/res_e/publications_e/world_tariff_profiles24_e.htm}{World Tariff Profiles 2024}), its members have agreed on strict bounds on tariffs to limit their flexibility, and in 2023 the trade-weighted average applied tariff on non-agricultural goods in major advanced economies was below 3\%. Moreover, tariffs are inherently unstable policy instruments that governments struggle to commit to. For instance, the sweeping 2025 U.S. \href{https://www.whitehouse.gov/presidential-actions/2025/04/regulating-imports-with-a-reciprocal-tariff-to-rectify-trade-practices-that-contribute-to-large-and-persistent-annual-united-states-goods-trade-deficits/}{``Liberation Day'' tariffs}, which imposed 10--50\% duties that escalated to a combined 145\% on Chinese goods within days, faced immediate legal challenges, market turbulence, and retaliatory spirals. The \href{https://www.whitehouse.gov/presidential-actions/2025/04/modifying-reciprocal-tariff-rates-to-reflect-trading-partner-retaliation-and-alignment/}{April 9 Executive Order} partially suspended these measures within a week, and courts later ruled that the tariffs exceeded presidential authority under IEEPA (\href{https://www.wilmerhale.com/en/insights/client-alerts/20250605-what-you-need-to-know-about-the-federal-court-decisions-threatening-the-trump-administrations-tariff-agenda}{IEEPA Tariff Litigation}).\label{motivating spatial competition}}; a labor union negotiating insider employment with a firm must account for the outsider labor supply (\cite{lisn2002});
a platform building its ecosystem may be restricted to contracting with major content creators, since users and low-quality creators can be numerous and transient, and thus harder to access; and a regulator monitoring corporate pollution must account for how boycotts by environmental activists affect product demand (\cite{frie2002}). These contracting situations differ in their externality structures. Competition between firms and among workers makes their actions strategic substitutes. For platforms, network effects (\cite{arms2006}; \cite{roti2003}) make users and producers complements in building the ecosystem, while fiercer competition can turn low-quality creators from complements into substitutes for high-quality ones. Finally, the strategic dependence between a polluting firm and activists is asymmetric because the firm's production stirs up boycotts, which in turn discourage pollution.

In these scenarios, a key concern facing insider agents and outsiders is \emph{strategic uncertainty}: in the presence of externalities, their actions vary with how they expect each other to behave, and this mutual skepticism may sustain multiple equilibria given their flexible strategic dependence. Although strategic uncertainty has been a main focus of the literature on contracting with externalities (e.g., \cite{sega2003}; \cite{wint2004}), previous studies make the important assumption that the principal can provide incentives for all payoff-relevant actions. In contrast, this paper considers a principal who relies on contracts that are not contingent on outsider actions to guarantee a unique outcome. Allowing for general externality structures, I ask the following research questions: What feature of the externality structure causes strategic uncertainty? What is the principal's optimal contract for robustly inducing a fixed outcome? How do non-contractibility and strategic uncertainty influence the optimal outcome and the distribution of surplus?

The model consists of a principal, an agent, and an outsider. The principal contracts with the agent {\`a} la \cite{sega2003} by offering a public bilateral contract specifying transfers that vary only with the agent's action. This contract is followed by a game in which the agent and the outsider act simultaneously. All players' payoffs may depend on the decisions of both the agent and the outsider, which allows for flexible externalities. To address strategic uncertainty, the principal maximizes her payoff over contracts that induce a unique mixed-strategy Nash equilibrium of this game. As is standard in contracting, her problem is decomposed into two steps: she designs the optimal contract for inducing each fixed outcome in the first step and searches for the optimal outcome in the second. My results rely mainly on one assumption: a one-dimensional measure of outsider decisions captures the incentives that each decision gives the agent to take higher actions. This assumption holds for a large family of agent payoffs, including those in all the above applications.\footnote{Section \ref{Assumptions} demonstrates that the setting includes the most widely studied externality structures, in which the actions of the agent and the outsider are pure complements or substitutes, as well as the case where the agent's payoff is separable in both actions. Section \ref{Applications} explains how applications with various externality structures are covered by the model.} This restriction makes the robust analysis tractable while preserving great flexibility in the players' strategic dependence, especially with little imposed on the outsider's payoff.

Without contracts contingent on outsider actions, strategic uncertainty emerges when the insider-outsider strategic dependence is \emph{symmetric in the target outcome} vis-{\`a}-vis the outside option, that is, when the outsider's best response to the target action, compared to his best response to the outside option, strengthens the agent's incentives to move toward the target.\footnote{The main model explicitly sets an outside option for the agent, whereas Section \ref{Discussions} argues that, under some conditions, the first-step characterization extends to a less restrictive setting in which the outside option is replaced by the unique (pure) Nash equilibrium played absent a contract.} To illustrate this point, consider cases with pure complements or substitutes, where the agent and the outsider always influence each other's incentives symmetrically. Here, an agent action above [below] the outside option moves the outsider's best response in the direction that raises [lowers] the agent's incentives, compared to the outsider's response to the outside option. For example, a domestic firm producing more reduces the market price and thus a foreign firm's production, which in turn raises the residual demand facing the domestic firm compared to the Cournot outcome. As a result, the agent's actions are self-reinforcing. To see how this causes a problem, imagine the benchmark in which the principal could select her favorite equilibrium. She would then design the contract conditional on the agent already expecting the outsider to respond to the target action. However, self-reinforcement leads to mutual skepticism: when the agent instead conjectures that the outsider reacts to the outside option, his incentives become insufficient for him to take the action recommended by the principal, because the transfer charged conditional on the target-outcome incentives is now too high. He then takes his outside option, which in turn justifies the outsider's reaction and sustains an undesired equilibrium. Consequently, without outsider contingency, strategic symmetry is the source of multiplicity. In particular, strategic uncertainty occurs even in submodular environments such as firm competition, where it would not arise if the principal could contract with both players (\cite{sega2003}).

To solve the principal's first-step problem of fully implementing a fixed outcome, I develop a duality approach to deal with general externality structures. This approach reformulates the principal's contract design problem as the choice of a series of agent expectations about outsider behavior, one for each agent action. The reformulation is natural because full implementation typically focuses on ruling out undesired expectations of the agent about non-contractible behavior.

Specifically, each fixed contract defines two dual problems. In the first problem, given an outsider decision, the agent chooses his optimal action from this contract. In the second problem, given an agent action, the principal selects the outsider's decision to maximize the transfer she can charge to sell this action in the same contract. Her optimal choice is dubbed the outsider's \emph{dual reply} to this action. I interpret a dual reply as the principal's ideal selection of how the agent expects the outsider to behave, because if the agent expected the outsider to play the dual reply to an action, he would be willing to pay the maximal transfer for it. Importantly, I show that the principal designs the optimal contract as if she selects the agent's expectations about outsider behavior, namely the dual replies, in a continuum of fictitious situations, one for each agent action between the outside option and the target outcome, subject to two constraints described below. Her objective is the sum of agent incentives across these situations, and in each situation she wants to raise agent incentives as much as possible. This objective is intuitive because agent incentives, measured by the marginal payoffs, capture the agent's willingness to pay for higher actions.

The key constraint on these expectation choices results from the principal's need to eliminate all undesired equilibria for full implementation. This constraint puts an upper bound on the agent incentives that each dual reply generates. Intuitively, at each action below the target, the principal cannot convince the agent to expect higher incentives than those generated by the outsider's best responses to the actions the principal has already conquered, that is, all actions between the outside option and this very action. To see why, suppose that she selects an expectation that generates even higher incentives and prices this action accordingly. A familiar mutual skepticism then arises: if the agent instead expects the outsider to respond to a lower action, his incentives fall, and the marginal transfer charged at this action appears too high. The agent therefore moves to a lower action, which in turn justifies the outsider's pessimistic reaction and sustains an undesired equilibrium.

The second constraint results from the principal's need to keep the target outcome an equilibrium. Hence, no expectation she selects can generate higher agent incentives than in the target outcome, since otherwise, as I show, the agent would profitably deviate from the target.

My first main result, Theorem \ref{main result}, characterizes the principal's maximal payoff guarantee and constructs a robustly optimal contract for inducing each fully implementable outcome, pure or mixed. This contract pushes the agent's incentives as high as the two constraints above allow by also offering intermediate actions, which lie between the outside option and the target action. Such an action is offered when the outsider's best response to it generates agent incentives at least as high as those rationalizable by any lower action, but no higher than in the target outcome. Accordingly, the marginal transfer is based on the \emph{cumulative maximal} agent incentives. Although these intermediate actions are not chosen on the equilibrium path, they serve to break undesired equilibria.\footnote{When the target is pure and the contract's transfers are slightly adjusted, the outsider's best reply to any undesired mixed strategy of the agent makes the agent want to deviate from the lowest action in this strategy to some higher action.} Intuitively, the principal uses intricate contractual terms to raise the transfers she can charge while robustly coordinating the outsider's action, but she sometimes leaves gaps in the contract to avoid either lowering these transfers or choking the desired outcome with excessive agent incentives.

I apply Theorem \ref{main result} to the regulation of international competition. If the regulator could contract with both firms through an industrial policy and a tariff, strategic uncertainty would not arise, so each optimal policy would contain a single plan (\cite{sega2003}). When tariffs have limited flexibility or are hard to put into effect, the robustly optimal industrial policy that I construct instead offers a continuum of plans. This contrast is consistent with policy practice, in which industrial policies are often more complex than tariffs.\footnote{For example, while the U.S. \href{https://www.congress.gov/bill/117th-congress/house-bill/5376/text}{Inflation Reduction Act} of 2022 involves intricate subsidy allocations and compliance checks for green technology, tariffs merely adjust border tax rates, as seen in Trump's 2025 flat 20--25\% levies on Chinese or Canadian goods.\label{industry and tariff}} My characterization also has a novel implication for how demand elasticities enter the design of industrial policies. If the regulator ignored strategic uncertainty or could use both policies, only the demand elasticity given the foreign firm's output in the target outcome would matter, whereas robustly optimal design depends on the elasticities along the foreign firm's best replies between the Cournot and target outcomes.

My second set of results, Propositions \ref{proposition-incentive attenuation} and \ref{proposition-integrated game}, uncovers critical features of the principal's second-step choice of the optimal outcome. Restricting attention to pure outcomes, I compare the optimal outcomes under robust and non-robust implementation to draw implications for how strategic uncertainty affects contracting output in the presence of a non-contractible outsider. Proposition \ref{proposition-incentive attenuation} shows that, for every feasible externality structure and principal payoff, the agent's incentives in the robustly optimal outcome are never stronger than without the robustness concern. This attenuation results from the strategic rents that the principal pays to uniquely induce an outcome. That is, she must compensate the agent for the payoffs he gives up when deviating from those outcomes in which his incentives are too low. To mitigate these costs, she steers the agent's expectations about outsider behavior, and thus his incentives, more conservatively than in the non-robust case, in which she can select these expectations directly and at no cost. Such an attenuation effect on agent incentives may bias the agent's action in either direction. For example, in regulating international competition, accounting for strategic uncertainty also attenuates domestic production because the regulator must pay extra subsidies to guarantee that the domestic firm is willing to produce enough to scare away foreign competition. On the other hand, a platform that cannot contract with low-quality imitators must pay extra rewards to guarantee high-quality production, given that content creators face network externalities. To reduce such strategic rents, the platform may induce the high-quality creators to over-produce, which invites more low-quality imitation and fiercer competition, thus attenuating the high-quality creators' incentives.

For a more precise second-step characterization, Proposition \ref{proposition-integrated game} further restricts attention to environments that are either supermodular or submodular, and describes the optimal outcomes through an \emph{integrated game}, in which the principal chooses the action in the agent's place and plays against the outsider. If the principal can both commit to a public contract and select the equilibrium, as in the non-robust case, she induces the Stackelberg outcome of this game, whereas the robustly optimal outcome must be a principal-preferred Nash outcome of the same game. In the integrated game, robustness thus turns the principal's commitment power into the power to select among Nash outcomes, which coordination power alone would give her; in this sense, the two powers are substitutes. Consequently, strategic uncertainty dampens principal surplus in two ways: the principal selects the optimal outcome as if she had no commitment power in the integrated game, and has to pay strategic rents to induce that outcome. When the principal's payoff depends only on the agent's action, the integrated game also describes private contracting: a principal who can commit to a public contract but faces a robustness concern then chooses among the same Nash outcomes as one who can select the equilibrium yet contracts privately, and the two powers are perfect substitutes when the Nash outcome is unique.\footnote{With several Nash outcomes, the two principals may still choose differently, since only the one who can commit pays strategic rents.}

Proposition \ref{proposition-integrated game} also offers a novel perspective on contractual privacy.\footnote{The privacy concept here is reminiscent of \cite{sega1999} rather than \cite{halr2021} or \cite{gali2025}. That is, I focus on whether the principal can commit to a contract observable to the outsider before he interacts with the agent.} Under public contracting, the principal induces her favorite Nash outcome (of the integrated game) while paying strategic rents. Under private contracting, she no longer pays the strategic rents, but when her payoff depends only on the agent's action, robustness leads her to expect the worst-case Nash outcome. Surprisingly, the principal may find private contracting strictly better, particularly when there is a unique Nash outcome: both regimes then induce it, but only public contracting costs her strategic rents. In sharp contrast, when the principal ignores strategic uncertainty, public contracting is never worse for her, since she now attains the Stackelberg outcome without paying any strategic rents. Consequently, accounting for strategic uncertainty can reverse the principal's preference over contractual privacy, leading to opposing predictions when, for example, a regulator, a labor union, or a platform can decide whether to keep its negotiations secret. This contrast highlights non-contractibility and strategic uncertainty as potential drivers of organizational practices.

\emph{Outline.} Section \ref{Model} sets up the model and discusses its assumptions and applications. Section \ref{Robustly Optimal Contract} first introduces the duality approach and shows that the principal designs the optimal contract as if she selects the agent's expectations about outsider behavior, subject to two constraints. It then presents the first-step characterization. Section \ref{Robustly Optimal Outcome} investigates the second-step problem in two special cases. Section \ref{Discussions} considers a setting in which the agent keeps full access to his actions even when he declines the contract. The Appendix contains all omitted proofs.

\paragraph{Literature review}
This paper belongs to a growing literature on contracting with externalities that focuses on unique implementation. One strand of this literature follows \cite{wint2004} and studies moral hazard problems with unobservable actions.\footnote{See also, e.g., \cite{elsp2015}; \cite{moya2020}; \cite{halr2021}; \cite{halr2022}; \cite{cugp2023}; \cite{moot2024}; \cite{capo2025}.} Another strand, pioneered by \cite{sega2003}, investigates settings with bilaterally contractible agent actions, as in this paper.\footnote{In this strand, \cite{bewi2012} examine a principal's optimal subsidies for inducing participation of heterogeneous agents (\cite{sast2012} study a related question, but their solution concept depends on inducing participation as a risk-dominant equilibrium); \cite{hakw2020} investigate optimal fundraising with endogenous externalities and asymmetric investors; \cite{ahls2022} and \cite{gali2024} study an information seller's optimal pricing and design of evidence structure; \cite{halr2024} find a monopolist's optimal pricing scheme for a network product; and \cite{chan2025} considers multi-agent contracting in a weighted potential environment.} My main departure from this literature lies in the principal's inability to cover all payoff-relevant decisions in the contract. Moreover, my duality approach handles flexible externality structures, not only the supermodular or submodular ones on which most of this literature focuses. \cite{gali2025} also study flexible externalities, which in their model arise from career concerns. Unlike the transfers here, which are deterministic but need not be linear, their transfers can be random, although each realization must be linear in the agent's action. Their agent's payoff is also linear.\footnote{In this paper's notation, the agent in \cite{gali2025} chooses his action as if his utility were $u_A(a,r)=ar+a\tau$, where $\tau$ is the incentive tool selected by the principal, so the transfers $t:=-a\tau$ must be linear in $a$. In that paper, however, $\tau$ is random and hidden from the outsider.} \cite{bewi2012} focus on strategic complements but also study flexible externalities in an extension. Their result, however, relies on a very different assumption: all agents can be divided into two groups, each with only positive or negative externalities among its members.

My results on what causes strategic uncertainty and how it affects the contracting outcome are closely related to \cite{sega2003}. Unlike Segal, who shows that complementarity is what causes multiplicity, I identify strategic symmetry as its source under partial contractual coverage, so multiplicity arises even with strategic substitutes. Furthermore, in \cite{sega2003}, strategic rents bias the principal's optimal choice of action outcome only downward. Here, they instead attenuate the agent's incentives, and this attenuation may bias his action in either direction, depending on the externality structure.

Finally, my focus on contractual privacy is related to \cite{sega1999}, \cite{halr2021}, and \cite{gali2025}. While \cite{halr2021} and \cite{gali2025} consider contracts that the principal commits to but that are hidden from third parties, my privacy concept is closer to that of \cite{sega1999}, who discusses whether the principal can commit. Segal's focus, however, differs from mine: he asks how privacy and transparency influence the effect of externalities on optimal contracting, whereas I ask how non-contractibility and strategic uncertainty change the principal's preferences over private and public contracting.

\section{Model}\label{Model}

A principal (she) contracts with an agent (he) before he interacts with an outsider to whom the principal cannot provide incentives. The agent can potentially access actions in $A:=[a_0,\overline{a}]$, whereas the outsider chooses his decision $r$ in $R:=[\underline{r},\overline{r}]$. An action outcome $(a,r)$ gives utility $u_A(a,r)$, $u_O(a,r)$, and $u_P(a,r)$ to the agent, the outsider, and the principal, respectively.

The main setting follows \cite{sega2003} and allows the principal to publicly offer the agent a bilateral contract that posts a menu of action-transfer plans. Specifically, a contract is $M=\{(a_j,t_j)_j\}\subset A\times\mathbb{R}$. The contract must contain the null plan $(a_0,0)$, which incorporates the agent's individual rationality, since he can always reject the contract and take the outside-option action $a_0$. In addition, I require that $M$ induce a compact graph of actions and transfers to ensure that the agent's optimal choice is well-defined. Let $\mathcal{M}$ collect all such contracts.
Here, the contract is bilateral, since each plan $(a_j,t_j)$ does not depend on the outsider's decision or include any transfer from the outsider.

\paragraph{Timing and payoffs} After the principal offers a contract $M$, both the agent and the outsider observe it. The two players then make simultaneous decisions: the outsider chooses his decision $r$ from $R$, and the agent picks one plan $(a,t)$ from $M$. Given the agent's choice, he plays the chosen action $a$ and pays the chosen transfer $t$ to the principal. The payoffs of the principal and the agent are thus $u_P(a,r)+t$ and $u_A(a,r)-t$, respectively. Unlike the contracting parties, the outsider makes no transfer and earns $u_O(a,r)$. Note that the transfers can be positive and negative, allowing the principal to both sell access ($t>0$) and provide incentives ($t<0$).

So far, the agent's access to actions is dictated by the principal, and the outside option is the lowest action. In Section \ref{Discussions}, I will argue that, with pure externalities and weakly concave agent utility, my main result extends to a setting in which, when the agent declines the principal's offer, he still has full access to $A$, and the unique agent-outsider Nash equilibrium is played, serving as the outside option.

\paragraph{Equilibrium and robust solution} Every contract $M$ induces a simultaneous-move game played by the agent and the outsider. A mixed-strategy Nash equilibrium consists of an agent strategy $\beta\in\Delta(M)$ and an outsider strategy $\rho\in\Delta(R)$, each satisfying the following optimality conditions
\begin{equation}
    \begin{aligned}
        (a,t)\in\supp(\beta)&\Rightarrow(a,t)\in\argmax_{(a',t')\in M}\int_{ R}[u_A(a',r)-t']d\rho(r); \\
        r\in\supp(\rho)&\Rightarrow r\in\argmax_{r'\in R}\int_Mu_O(a,r')d\beta(a,t).
    \end{aligned}
\end{equation}
To discuss how the principal tackles strategic uncertainty by inducing play robustly, I let $\mathcal{M}_U$ collect all contracts that induce a unique such equilibrium. For every $M\in\mathcal{M}_U$, denote the induced equilibrium strategy profile by $(\beta_M,\rho_M)$. Note that every agent action $a\in A$ (and the outsider's best response to it, which is unique under the strict concavity of $u_O$ assumed in Section \ref{Assumptions}) can be uniquely induced by offering $(a,t)$ with a sufficiently low $t$.

This paper investigates how the principal maximizes her expected payoff in the unique equilibrium that her contract induces. In particular, the principal solves the following problem

\begin{definition}\label{MRG-def}
    \emph{The principal's} payoff guarantee \emph{given contract $M\in\mathcal{M}_U$ is}
    \begin{equation}\label{U_P(M)}
        U_P(M):=\int_M\int_{ R}[u_P(a,r)+t]d\beta_M(a,t)d\rho_M(r).
    \end{equation}
    \emph{Her }maximal payoff guarantee \emph{is}
    \begin{equation}
        U_P^*:=\sup_{M\in\mathcal{M}_U}U_P(M).
    \end{equation}
    \emph{A contract $M^*$ is }robustly optimal\emph{ if there is a sequence of contracts $(M_n)_{n=1}^{\infty}$ such that
    \begin{enumerate}[nolistsep]
        \item the sequence $(M_n)_{n=1}^{\infty}\subset\mathcal{M}_U$ and it converges to $M^*$ with respect to the Hausdorff metric;
        \item the sequence of payoff guarantees $\left(U_P(M_n)\right)_{n=1}^{\infty}$ converges to $U_P^*$.
    \end{enumerate}}
\end{definition}

\subsection{Assumptions}\label{Assumptions}
To ease exposition, this paper considers twice-differentiable payoff functions $u_A$, $u_O$, and $u_P$.
Moreover, I assume that
the outsider's utility function $u_O$ is strictly concave in his action $r$. Hence, the outsider's best response to every agent strategy $\beta$ is unique, and I denote this best-response mapping as $r(\beta)$. Note that even though the strategy $\beta$ selects action-transfer plans, the outsider's unique response depends only on how the agent chooses his action. So, for every mixed action $\alpha\in\Delta(A)$, I let the associated outsider response be $r(\alpha)$. Moreover, the outsider's best response to a pure action $a$ is simply denoted as $r(a)$. Throughout the paper, the above assumptions are maintained.

The uniqueness of the outsider's best response allows me to solve the principal's problem in two steps: first, I fix an action outcome $(\alpha,r(\alpha))$, where $\alpha\in\Delta(A)$ is a potentially mixed decision of the agent and $r(\alpha)$ the associated outsider reply, and find the robustly optimal contract for fully implementing this action outcome; second, I search for the optimal outcome conditional on the first-step solution. To define the first-step solution, I let $\mathcal{M}_U(\alpha)$ collect all contracts that uniquely induce $(\alpha,r(\alpha))$. Note that not every outcome is implementable since $\mathcal{M}_U(\alpha)$ may be empty. I will discuss implementability in Section \ref{Robustly Optimal Contract}, and for now I let $\mathcal{A}_{FI}$ be the set of mixed actions that are fully implementable. For all $\alpha\in\mathcal{A}_{FI}$, the \emph{maximal payoff guarantee of inducing $\alpha$} is
\begin{equation}\label{first step}
    U_P^*(\alpha):=\sup_{M\in\mathcal{M}_U(\alpha)}U_P(M).
\end{equation}
A contract $M^*(\alpha)$ is \emph{robustly optimal for inducing $\alpha$} if a sequence $(M_n)_{n=1}^{\infty}\subset\mathcal{M}_U(\alpha)$ converges to $M^*(\alpha)$ and its payoff guarantees converge to $U_P^*(\alpha)$ in the sense of Definition \ref{MRG-def}. Hence, the second-step problem is $U_P^*=\sup_{\alpha\in\mathcal{A}_{FI}}U_P^*(\alpha)$. \\

My results depend on two additional assumptions, Assumptions \ref{ranked incentives} and \ref{intuitive mixture}, that generalize the setting beyond pure externalities, which may be the most widely studied family of strategic structures. In particular, pure externalities include environments where players' decisions are either supermodular (strategic complements) or submodular (strategic substitutes). Following the terminology of \cite{sega1999,sega2003}, I define increasing and decreasing externalities for this setting, which together constitute the category of pure externalities

\begin{definition}
    \emph{Externalities are }increasing \emph{[}decreasing\emph{] if $\frac{\partial^2u_A}{\partial a\partial r}\geq0$ and $\frac{\partial^2u_O}{\partial a\partial r}\geq0$ [$\frac{\partial^2u_A}{\partial a\partial r}\leq0$ and $\frac{\partial^2u_O}{\partial a\partial r}\leq0$] for all $(a,r)$. Moreover, externalities are }pure\emph{ if they are either increasing or decreasing.}
\end{definition}

With increasing [decreasing] externalities, each player's selection of a higher action enhances [reduces] the opponent's marginal payoff from choosing a higher action. Given the twice-differentiability assumption, the definitions here coincide with those of Segal, which are based on the payoff functions having increasing [decreasing] differences. Therefore, pure externalities are the leading case of settings where players have \emph{symmetric strategic dependence}.

While most discussions on contracting with externalities restrict attention to this pure family, this paper extends the scope by allowing for mixed externalities. To generalize, I introduce two assumptions below, which I also maintain throughout. Pure externalities satisfy the second assumption, and also the first if $\frac{\partial^2u_A}{\partial a\partial r}\neq0$ everywhere.

\paragraph{Ranked incentives}
The first assumption I impose demands that the agent have ranked incentives. Specifically, I say one outsider action $r_1$ \emph{weakly dominates} another $r_2$ \emph{in the order of agent incentives}, denoted $r_1\succsim_{AI}r_2$, if $\frac{\partial u_A}{\partial a}(a,r_1)\geq\frac{\partial u_A}{\partial a}(a,r_2)$ for all $a$. In other words, $r_1\succsim_{AI}r_2$ if the agent's marginal gain from choosing a higher action is weakly greater given the opponent's decision $r_1$ than given $r_2$. As one can see, this order is transitive. Its strict and indifference parts are denoted $\succ_{AI}$ and $\sim_{AI}$. The assumption below then requires that this order also be complete, and that the agent's incentives increase strictly in it

\begin{assumption}\label{ranked incentives}
    Ranked incentives: \emph{$\succsim_{AI}$ is a complete order over $R$, and $r_1\succ_{AI}r_2$ implies $\frac{\partial u_A}{\partial a}(a,r_1)>\frac{\partial u_A}{\partial a}(a,r_2)$ for all $a$.}
\end{assumption}

The leading example, pure externalities, has ranked agent incentives when $\frac{\partial^2u_A}{\partial a\partial r}\neq0$ everywhere, because increasing [decreasing] externalities then make $\succsim_{AI}$ equivalent to the increasing [decreasing] order of outsider actions. Another case of ranked incentives that matches applications with non-pure externalities (see Section \ref{Applications}) is separable payoffs. In particular, when the agent's payoff function takes the form $u_A(a,r)=g(a)h(r)+k(a)$ with $g'>0$, $r_1\succsim_{AI}r_2$ if and only if $h(r_1)\geq h(r_2)$.

By the representation theorem of \cite{debr1954}, when the transitive order $\succsim_{AI}$ is complete (ranked agent incentives) and continuous (twice-differentiability of $u_A$), it is represented by a continuous ``utility'' function. For example, fixing any $a$, the marginal utility $\frac{\partial u_A}{\partial a}(a,\cdot)$ is one such representation. Therefore, I am able to define the maximum and minimum for this order as $\max_{AI}$ and $\min_{AI}$. If multiple outsider actions are equivalent in this order, I simply use one of them to represent the equivalence class. Also, I will frequently use this order to compare sets of outsider actions, say $S\succsim_{AI}S'$, which means that \emph{every} outsider action in $S$ generates weakly higher agent incentives than \emph{every}thing in $S'$. Likewise, a comparison between a set and a single action means that the relation holds for all elements of the set.

\paragraph{Intuitive mixture} The second nice property of pure externalities is that the order of agent incentives $\succsim_{AI}$ is quasi-concave, which I formalize as the following condition
\begin{assumption}\label{intuitive mixture}
    Intuitive mixture: \emph{for all $r_1$, $r_2$, and $r\in[r_1,r_2]$, $r\succsim_{AI}\min_{AI}\{r_1,r_2\}$.}
\end{assumption}
In other words, the assumption requires that if $r(a_1)\precsim_{AI}r(a_2)$, then the outsider's best reply to any mixture of $a_1$ and $a_2$ must satisfy $r(a_1)\precsim_{AI}r(\text{mixing $a_1$ and $a_2$})$. Notice that the best replies to such mixtures are exactly the outsider actions in $[r(a_1),r(a_2)]$. Assumption \ref{intuitive mixture} thus has an intuitive interpretation: the agent's incentives are improved by mixing with actions that induce greater agent incentives. The case with pure externalities satisfies this condition because $\succsim_{AI}$ now coincides with a monotonic order, which must be quasi-concave. Assumption \ref{intuitive mixture} is important for the proof of Theorem \ref{main result}. In the contracts constructed there, it ensures that whenever the agent mixes over actions below the lowest target action, the outsider's best reply still makes the agent want to deviate upward from the lowest of these actions, so no such mixture is played in equilibrium.

Although Assumptions \ref{ranked incentives} and \ref{intuitive mixture} impose structure on the agent's payoff, I have not restricted the outsider's behavior beyond the strict concavity of $u_O$, which makes his best reply $r(\cdot)$ unique and continuous but otherwise unrestricted. This allows the analysis to cover a wide range of applications, even those with arbitrary mixed externalities. Consider the above case of separable payoffs for example. Here, I simply assume that $h(\cdot)$ is quasi-concave, while the arbitrary form of $r(\cdot)$ can generate complex strategic interaction.

\subsection{Applications}\label{Applications}

This section provides several motivating examples of the model. As will be shown, they match the main setup, except that spatial competition matches the full-access variant of this setup. Those with explicit payoffs satisfy the assumptions I have made so far. The applications cover settings with both pure and non-pure externalities, and more details will be provided for those cases with strategic substitutes and mixed strategic dependence because later on they will be revisited to demonstrate the novelty of the results.

\paragraph{Spatial competition} The agent and the outsider are firms competing across jurisdictions in a duopoly Cournot game. The two jurisdictions can be, for instance, different countries or states. The principal is the government of one jurisdiction and can commit to industrial policies that incentivize production, but only that of the domestic firm (agent). Multiple reasons may prevent the government from contracting on foreign firms' (outsider's) decision: it lacks the sovereignty to impose any incentive plans on foreigners; intervening in international trade violates established trade rules (e.g., WTO trade rules); the use of trade policies such as tariffs is usually strategic and temporary, serving as tools for negotiation and retaliation, instead of permanent terms that governments will commit to (recall Footnote \ref{motivating spatial competition}). As one can easily see, the competitive nature of the firms' interaction gives rise to decreasing externalities.

I formalize this Cournot environment. The domestic firm's payoff function is $u_A(a,r)=aP(a,r)-c_A(a)$ and the foreign firm's payoff is $u_O(a,r)=rP(a,r)-c_O(r)$, where $a$ and $r$ are their respective production levels, $P$ is the inverse demand function, and $c_A$ and $c_O$ are the cost functions facing each of them. To fix ideas, I will work with the simple specification with $P(a,r)=1+c-a-r$, $c_A(a)=ca$, and $c_O(r)=cr$. This then yields strategic substitutes $\frac{\partial^2u_A}{\partial a\partial r}=\frac{\partial^2u_O}{\partial a\partial r}=-1<0$.  Though the government's objective can vary, I focus on two common goals: the government may care about social efficiency and its payoff is $u_P(a,r)=\int_0^{a+r}(1-q)dq$, so the preferred outcome induces higher total production than the Nash level; by contrast, the government can also be a regulator of emissions that aims to reduce the total production and has a payoff function $u_P(a,r)=g(a+r)$ with a strictly decreasing $g$.

One can already see how the above example differs from the main setting: the agent has full access to all actions even when he declines the contract (i.e., the domestic firm can refuse to claim the subsidies specified in the industrial policy). I discuss this full-access scenario in Section \ref{Discussions} and argue that the first-step characterization can be readily extended to this application, which has pure externalities and weakly concave agent utility. The outside option is no longer the lowest action, but rather the Nash outcome (in this example, $a_0=r(a_0)=\frac{1}{3}$). Thus, for full implementation, the contracting problem of the principal here coincides with that of the main setting with outside option $a_0$, except that the principal must target an action higher than $a_0$ and must use non-positive transfers. To induce a target action lower than the Nash point, simply replace $a$ with $-a$, which still preserves all the assumptions because pure externalities are still pure.

\paragraph{Union contracting} The principal is a labor union that contracts with a firm (agent) in the presence of non-union workers (outsider). The union negotiates with the firm on the number of unionized workers $a$ it hires and the total wage payment $t$ it pays to them.
However, the firm can also hire from a fringe labor supply not covered by the union, including independent contractors, gig workers, or employees searching on the job. This fringe labor supply and the wage it faces are simultaneously determined in a general equilibrium. Note that decreasing externalities emerge in this application because union supply and fringe supply are substitutes.

In particular, the union proposes a menu that contains plans of the form $(a,t)$. The firm earns revenue $f(a+r)$ from $a$ union workers and $r$ fringe workers, with $f''<0$. Fringe workers are paid the wage $w_f(r)$, where $w_f$ is the nondecreasing inverse supply curve of fringe labor, so the firm's payoff net of fringe wages is $u_A(a,r)=f(a+r)-w_f(r)r$. Given $a$, a competitive fringe market employs the fringe labor $r$ at which the firm's marginal revenue equals the fringe wage, that is, $f'(a+r)=w_f(r)$ at an interior solution. Since this employment level maximizes the surplus of the fringe market, I model the outsider's payoff as this surplus, $u_O(a,r)=f(a+r)-f(a)-\int_{0}^{r}w_f(s)ds$, which is strictly concave in $r$. Both $\frac{\partial^2u_A}{\partial a\partial r}$ and $\frac{\partial^2u_O}{\partial a\partial r}$ equal $f''(a+r)<0$, so externalities are decreasing. In words, more union workers lower the firm's marginal revenue and thus discourage fringe employment, so $r(\cdot)$ is decreasing, while more fringe workers lower the firm's marginal gain from hiring union workers. This establishes the submodular nature of the interaction between the firm and the non-union outsiders.

\paragraph{Two-sided market} The agent and the outsider are two sides of a market (such as producers and consumers) who interact via a platform (principal). As pointed out by, for instance, \cite{roti2003} and \cite{arms2006}, such markets usually manifest network externalities between the two sides. However, it is sometimes impractical for the platform to provide incentives to one side of the market. One salient example is that while e-commerce marketplaces and social media provide extensive monetization plans to sellers and content creators, these platforms usually refrain from contracting with an enormous number of users due to high transaction costs and the users' transient nature. As is well known, network effects constitute a hallmark form of increasing externalities. \\

In addition to the above applications, I introduce another two examples in which the externality structure may not be pure. The first example below incorporates asymmetric strategic dependence while the second one allows for partially pure externalities.

\paragraph{Regulation with boycott} The agent is a monopoly firm that chooses a quantity to be sold in a market. However, the outsider is a group of environmental activists who may boycott the firm's production to fight against emissions. The principal is a government that cares about social efficiency but internalizes the negative externalities of emissions. In particular, the scale of the boycott (outsider action) has negative externalities on the consumers' willingness to pay for the firm's product, whereas the firm's production scale (agent action) facilitates boycotting as a collective decision. In this example, the strategic interdependence does not demonstrate pure externalities but is instead asymmetric. Since a larger boycott lowers the firm's marginal revenue at every quantity, the actions of the outsider are completely ranked in the order of agent incentives, which coincides with the decreasing order of outsider actions.

\paragraph{Networked competition} The principal is a two-sided platform. The agent is the group of major high-quality content creators to whom the platform can provide incentives through monetization plans. On the other hand, the outsider consists of transient low-quality content creators who are hardly accessible to the platform. The interaction of these two players presents both the competition effect and the network effect. Crucially, these two effects are heterogeneous across groups: high-quality creators are able to produce innovative content, generating great spillovers, so low-quality creators can benefit from imitation; however, the spillovers from the other side are often limited. Specifically, I consider utility functions $u_A(a,r)=g_A(a)(\beta_Or-r^2)-c_A(a)$ and $u_O(a,r)=g_O(r)(\beta_Aa-a^2)-c_O(r)$. Here, $a$ and $r$ correspond to the production levels of both players, $\beta_O$ and $\beta_A$ represent the scales of their spillovers to the other side, and the quadratic terms capture the competition effects. I assume that $g_A'>0$ and $g_O'>0$, and that $g_O$ is concave and $c_O$ is strictly convex, so that $u_O$ is strictly concave in $r$. To match the heterogeneous network effects, I consider $\beta_A\gg\beta_{O}=1$, in the sense that $\overline{a}<\frac{\beta_A}{2}$. As a result, high-quality production always exerts increasing externalities on the low-quality sector, whereas the low-quality content is initially a strategic complement to the high-quality innovation, since it can, for example, attract more users. However, as competition becomes fiercer, the high-quality sector faces decreasing externalities from low-quality imitation. In this application, pure externalities appear only when the outsider exerts a low effort.

\section{Optimal Implementation}\label{Robustly Optimal Contract}

This section provides the first-step characterization. In particular, I take any fully implementable mixed action of the agent $\alpha\in\mathcal{A}_{FI}$, and determine a robustly optimal contract for inducing it. In Section \ref{A Duality Approach to General Characterization}, I begin by introducing a duality approach, which is useful for dealing with the general payoff structure and the full implementation of mixed-strategy equilibria. Section \ref{Main Result} presents the main result, Theorem \ref{main result}, and illustrates the principal's robustly optimal strategy with a few examples.

\subsection{A Duality Approach to General Characterization}\label{A Duality Approach to General Characterization}

To start with, I take any contract $M\in\mathcal{M}$ and define two problems: for all $a\in A$ and $r\in R$,
\begin{align}
    V(r;M)&:=\max_{(a',t')\in M}u_A(a',r)-t';\label{V(r;M)} \\
    T(a;M)&:=\max_{r'\in R}u_A(a,r')-V(r';M),\text{ and its solution set is }R(a;M).\label{T(a;M)}
\end{align}
The above two problems are dual in the following sense. On the one hand, (\ref{V(r;M)}) refers to the agent's optimal plan choice given outsider action $r$ and contract $M$, with his value denoted $V(r;M)$. On the other hand, (\ref{T(a;M)}) can be viewed as a transfer maximization problem of the principal in response to the agent's optimal plan choice. Specifically, suppose that the principal, having already offered contract $M$, wants to induce action $a$. Then $T(a;M)$ is the maximal price she can charge when the outsider's action is also within her discretion. In other words, for each optimal reply $r'\in R(a;M)$ in problem (\ref{T(a;M)}), $u_A(a,r')-V(r';M)$ is the agent's willingness to pay for $a$ since $V(r';M)$ represents what he would gain by choosing otherwise. Therefore, if the principal adds a new plan $(a,T(a;M))$ to $M$, the agent is willing to choose this plan when he expects the outsider to play $r'$. I thus call $T(\cdot;M)$ the \emph{dual transfer} function induced by contract $M$, and the solution correspondence $R(\cdot;M)$ contains the outsider's \emph{dual replies} to agent decisions.

I am interested in this duality form because it has a few nice properties. First, if some plan $(a,t)\in M$, then the dual transfer satisfies $T(a;M)\leq t$. This is because the agent will never pick action $a$ at a transfer above $t$ given that $(a,t)$ is offered. Moreover, a contract $M$ induces an equilibrium with the outcome $(a,r(a))$ if and only if $(a,t)\in M$, $T(a;M)=t$, and $r(a)\in R(a;M)$. The reason is straightforward: the dual replies $R(a;M)$ maintain $(a,T(a;M))$ as the agent's optimal choice given $M$, and $T(a;M)=t$ means that this plan is already offered in $M$. A related property is that if $r(a)\succ_{AI} R(a;M)$, then given that the outsider expects the agent to play $a$ and thus plays $r(a)$, the agent must now want to deviate to some higher action offered in $M$. This is because $r(a)$ generates higher agent incentives than any action in $R(a;M)$, and an outsider action in this set is needed to sustain $a$ in equilibrium. Similarly, $r(a)\prec_{AI}R(a;M)$ indicates a downward deviation in the candidate equilibrium where the agent would play $a$.

As a result, one feature of each fully implementing contract is that every transfer chosen on path in the unique equilibrium must coincide with the dual transfer. Claim \ref{claim 1} formalizes this feature

\begin{claim}\label{claim 1}
    For each $M$ that fully implements $\alpha$, $U_P(M)=\int_A\left[u_P(a,r(\alpha))+T(a;M)\right]d\alpha(a)$.
\end{claim}

One can see that the principal's payoff guarantee $U_P(M)$ is rewritten in the duality form, by simply replacing the on-path transfers with the dual transfers in its definition (\ref{U_P(M)}). For the first-step implementation, the first part $\int_{A}u_P(a,r(\alpha))d\alpha(a)$ is irrelevant because it is determined by the target outcome $\alpha$ alone, so contract design matters only through dual transfers.

Claim \ref{claim 2} reveals how convenient this form is. In particular, I show that the dual transfer function satisfies an integral envelope theorem. By definition (\ref{T(a;M)}), the envelope theorem (if applicable at $a$) suggests that $\frac{\partial}{\partial a}T(a;M)=\frac{\partial}{\partial a}u_A(a,\widetilde{r}(a))$ with $\widetilde{r}(a)\in R(a;M)$. Claim \ref{claim 2} then demonstrates that this derivative form holds almost everywhere and the dual transfer is absolutely continuous

\begin{claim}\label{claim 2}
    For all $M$ and $a$, $T(a;M)=T(a_0;M)+\int_{a_0}^a\frac{\partial}{\partial a}u_A(a',\widetilde{r}(a'))da'$ with each $\widetilde{r}(a')\in R(a';M)$.
\end{claim}

Notably, Claims \ref{claim 1} and \ref{claim 2} together reformulate the principal's first-step problem (\ref{first step}) as one in which she selects a series of decisions $\widetilde{r}$ for the outsider in the dual problem (\ref{T(a;M)}). Note that each selection $\widetilde{r}(a')$ enters the principal's objective via the agent's local incentive $\frac{\partial}{\partial a}u_A(a',\widetilde{r}(a'))$. As an interpretation, the principal can be seen as selecting the agent's expectation about outsider behavior $\widetilde{r}(a')$ conditional on the agent ``currently playing'' $a'$, and this derivative $\frac{\partial}{\partial a}u_A(a',\widetilde{r}(a'))$ determines the agent's willingness to pay for a higher action, also reflecting the local transfer increment that the principal can charge.

Suppose that the principal does not face any constraint. Claim \ref{claim 2} then indicates that she would set every $\widetilde{r}$ to be the outsider action generating the greatest agent incentives. However, the principal faces two constraints. First, she faces the \emph{equilibrium-keeping constraint}: she must ensure that an equilibrium with the desired outcome $(\alpha,r(\alpha))$ exists. Second, she also deals with the \emph{equilibrium-breaking constraint}: for full implementation, she needs to eliminate any other unwanted equilibrium.

To formalize these two constraints, I denote the highest and lowest on-path actions of $\alpha$ as $\overline{a}(\alpha):=\max\supp(\alpha)$ and $\underline{a}(\alpha):=\min\supp(\alpha)$, respectively. Claim \ref{claim 3} then relates to the equilibrium-keeping constraint above and imposes an upper bound on the outsider actions that the principal can pick 

\begin{claim}\label{claim 3}
    For each $M$ that fully implements $\alpha$, every $a<\overline{a}(\alpha)$ satisfies $R(a;M)\precsim_{AI}r(\alpha)$.
\end{claim}

To understand why $r(\alpha)$ appears here, note that it is the outsider's choice in the desired equilibrium. In fact, if the principal could completely ignore strategic uncertainty and choose her favorite equilibrium, that is, partially implement outcome $\alpha$, then she would set $\widetilde{r}=r(\alpha)$ in Claim \ref{claim 2}. This is because, for partial implementation, the agent's actions are priced conditional on the outsider already choosing $r(\alpha)$.\footnote{For the partial implementation of $\alpha$, the principal simply offers the on-path actions, with a transfer $t=u_A(a,r(\alpha))-u_A(a_0,r(\alpha))$ for each action $a$. In this case, the principal gains the total principal-agent welfare net of the agent's outside value conditional on the on-path reply: $\int_{\underline{a}(\alpha)}^{\overline{a}(\alpha)}\left[u_P(a,r(\alpha))+u_A(a,r(\alpha))\right]d\alpha(a)-u_A(a_0,r(\alpha))=U_0(\alpha)+\int_{a_0}^{\underline{a}(\alpha)}\frac{\partial}{\partial a}u_A(a,r(\alpha))da$, where $U_0(\cdot)$ is given by (\ref{upper bound}). I intentionally write this partial-implementation value in this form to compare it with the robust value provided in Proposition \ref{proposition-upper bound}.\label{partial implementation}} Such an upper bound thus represents the need to maintain the desired equilibrium.

The proof of Claim \ref{claim 3} relies on showing that the dual replies are nondecreasing in the order of agent incentives, that is, for all $a_2>a_1$, one has $R(a_2;M)\succsim_{AI}R(a_1;M)$. Claim \ref{claim 3} then results from $r(\alpha)\in R(\overline{a}(\alpha);M)$ because $r(\alpha)$ and $\overline{a}(\alpha)$ are both equilibrium actions.

Crucially, the key lemma that guides the main result is Claim \ref{claim 4}, which comes from the equilibrium-breaking constraint, under which the principal must ensure that no outcomes other than $\alpha$ appear in equilibrium. To state the result, I define the principal's \emph{cumulative optimal reply}
\[\overline{r}(a):=\max_{AI}\{r(a'):a'\in[a_0,a]\},\footnote{Although this problem may admit multiple solutions, recall that each outsider action denotes the entire equivalence class.}\]
which is, among the outsider actions that can be rationalized by agent decisions no higher than $a$, the one generating the greatest agent incentives. I then show that this cumulative peak forms another upper bound

\begin{claim}\label{claim 4}
    For each $M$ that fully implements $\alpha$, every $a<\underline{a}(\alpha)$ satisfies $R(a;M)\precsim_{AI}\overline{r}(a)$.
\end{claim}

Recall that I interpreted the principal's choice of $\widetilde{r}$ as picking the agent's expectation about outsider behavior. Therefore, the above condition $R(a;M)\precsim_{AI}\overline{r}(a)$ dictates an intuitive rule that the principal must follow for full implementation. That is, she cannot convince the agent to expect higher incentives than those generated by the outsider's best responses to the agent actions that the principal has already ``conquered''. Specifically, at each $a$, the principal has eliminated $[a_0,a)$ from equilibrium, so her pricing conditional on the outsider's best reply to any of these actions will still induce the agent to deviate from $a$. In contrast, suppose that the principal designs the marginal transfer conditional on the outsider reacting to some $a'>a$ with $r(a')\succ_{AI}\overline{r}(a)$. Such a reaction produces very high agent incentives, resulting in a relatively high marginal transfer. As a result, a mutual skepticism problem arises here: if the agent instead believes that the outsider does not expect his action $a'$ as recommended by the principal, the high transfer tends to push him back to a lower action, which may sustain the outsider's low expectation about his action in a bad equilibrium, violating robustness. As I explain in the following, this logic guides the proof.

The proof of Claim \ref{claim 4} is involved, but the main idea is to first suppose that some $a'$ and $r\in R(a';M)$ exist such that $r\succ_{AI}\overline{r}(a')$, and then to construct an undesirable equilibrium to obtain a contradiction to full implementation. Supposing that $r\succ_{AI}\overline{r}(a')$, I know that $r\succ_{AI}r(a)$ for all $a\in[a_0,a']$, with which I manage to show the existence of a plan $(a,t)$ that is offered in $M$ and satisfies $r(a)\prec_{AI}R(a;M)$. A central step is to show that this plan must lead to the existence of an equilibrium where the agent's strategy is bounded above by $a$ (so this is a bad equilibrium). Intuitively, this intermediate step reveals that the principal's elimination of bad outcomes must be directional. Specifically, it imposes $r(a)\succ_{AI}R(a;M)$ on every action $a<\underline{a}(\alpha)$ that the agent would choose under some outsider decision. This, as I have argued previously, indicates that the agent now wants to deviate to a higher action given the outsider's best response to $a$. Thus, full implementation necessarily breaks bad outcomes at these actions by inducing \emph{upward} deviations alone.

Combining Claims \ref{claim 1}--\ref{claim 4} yields an upper bound for the principal's payoff guaranteed by each $M$ that fully implements $\alpha$. It will become clear that this bound is independent of $M$, so it also bounds the maximal payoff guarantee of inducing $\alpha$. In particular, one obtains this bound by maximally raising the agent's incentives subject to Claims \ref{claim 3} and \ref{claim 4}; that is, the principal selects $\widetilde{r}(a)=r^*_{\alpha}(a)$ with $r^*_{\alpha}(a):=\min_{AI}\{\overline{r}(a),r(\alpha)\}$ for all $a<\underline{a}(\alpha)$, and $r^*_{\alpha}(a):=r(\alpha)$ for all $a\in[\underline{a}(\alpha),\overline{a}(\alpha)]$. As a result, I obtain the following proposition

\begin{proposition}\label{proposition-upper bound}
    The maximal payoff guarantee of inducing $\alpha$ is bounded above as follows
    \begin{equation}\label{upper bound}
        \begin{aligned}
            U_P^*(\alpha)&\leq U_0(\alpha)+\overline{T}(\alpha)\text{, where} \\
            U_0(\alpha)&:=\int_{\underline{a}(\alpha)}^{\overline{a}(\alpha)}\left[u_P(a,r(\alpha))+u_A(a,r(\alpha))\right]d\alpha(a)-u_A(\underline{a}(\alpha),r(\alpha));\text{ and} \\
            \overline{T}(\alpha)&:=\int_{a_0}^{\underline{a}(\alpha)}\frac{\partial}{\partial a}u_A(a,r^*_{\alpha}(a))da=\int_{a_0}^{\underline{a}(\alpha)}\min\left\{\frac{\partial}{\partial a}u_A(a,\overline{r}(a)),\frac{\partial}{\partial a}u_A(a,r(\alpha))\right\}da.
        \end{aligned}
    \end{equation}
\end{proposition}

Note that the above payoff bound (\ref{upper bound}) consists of two parts. The first part, $U_0(\alpha)$, equals the total principal-agent welfare in the induced outcome $\int(u_P+u_A)d\alpha$, net of the agent's gain at his lowest on-path choice $\underline{a}(\alpha)$. This part can be roughly viewed as the principal's direct payoff plus the agent's willingness to pay for $\alpha$ vis-{\`a}-vis his ``base'' action $\underline{a}(\alpha)$. In the specific case where $\alpha$ is a degenerate action $a$, one simply has $U_0(a)=u_P(a,r(a))$. Moreover, the second part, $\overline{T}(\alpha)$, corresponds to the maximal transfer that the principal can charge for robustly inducing the ``base'' action $\underline{a}(\alpha)$. The derivative of this transfer is bounded by the cumulative peak incentives under $\overline{r}(a)$ (as a result of the need to break every bad outcome) and the on-path incentives under $r(\alpha)$ (as a result of the need to maintain the good outcome).

In fact, Proposition \ref{proposition-upper bound} provides a tight bound, as the main result will soon construct fully implementing contracts with payoff guarantees arbitrarily close to (\ref{upper bound}). In this way, I also complete the proof of the main result. Note that the principal's full-implementation value (\ref{upper bound}) becomes that of partial implementation (see Footnote \ref{partial implementation}) if the integrand $\min\left\{\frac{\partial}{\partial a}u_A(a,\overline{r}(a)),\frac{\partial}{\partial a}u_A(a,r(\alpha))\right\}$ is replaced with a weakly larger term $\frac{\partial}{\partial a}u_A(a,r(\alpha))$.

\subsection{Main Result: First-Step Implementation}\label{Main Result}

In this section, I fix a mixed action $\alpha\in\mathcal{A}_{FI}$ of the agent and explicitly construct a contract that attains the payoff upper bound (\ref{upper bound}) as the limit of fully implementing contracts. Later, I provide the necessary and sufficient conditions for the full implementability of each $\alpha$. To build this contract, I first take the maximal selection of outsider decisions, denoted $r^*_{\alpha}$, that I adopted to construct (\ref{upper bound}): $r^*_{\alpha}(a)=\min_{AI}\{\overline{r}(a),r(\alpha)\}$ for all $a<\underline{a}(\alpha)$; and $r^*_{\alpha}(a)=r(\alpha)$ for all $a\in[\underline{a}(\alpha),\overline{a}(\alpha)]$. With this selection, I construct the desired contract
\begin{equation}\label{optimal contract}
    \begin{aligned}
        A^*_{\alpha}&:=\{a_0\}\cup{}\{a<\underline{a}(\alpha):r(a)\sim_{AI}r^*_{\alpha}(a)\}\cup\supp(\alpha); \\
        t^*_{\alpha}(a)&:=\int_{a_0}^a\frac{\partial}{\partial a}u_A(a',r^*_{\alpha}(a'))da'\text{, for all }a\in A^*_{\alpha}.
    \end{aligned}
\end{equation}
The following main result establishes the robust optimality of this contract

\begin{theorem}\label{main result}
    A robustly optimal contract for inducing $\alpha\in\mathcal{A}_{FI}$ is given by $M^*(\alpha)=\{(a,t^*_{\alpha}(a)):a\in A^*_{\alpha}\}$, and the maximal payoff guarantee of inducing $\alpha$ is $U_P^*(\alpha)=U_0(\alpha)+\overline{T}(\alpha)$.
\end{theorem}

The key feature of $M^*(\alpha)$ is that it uses intermediate plans as strategic baits to steer outsider behavior and agent incentive in a ``maximal'' manner. To see this, observe that even though only those plans that offer the actions in $\supp(\alpha)$ are chosen on the path of the desired equilibrium, many others with actions in $[a_0,\underline{a}(\alpha))$ are also present, but never selected. These off-path options are introduced to destabilize undesired outcomes. In particular, the transfers are constructed in such a way that, once the marginal transfer at each off-path action is lowered slightly, for each agent strategy $\beta$ that assigns a positive probability to some of these off-path plans, there will always be a plan that gives the agent a strictly higher payoff than $\beta$'s lowest-action choice or than its choice of $\overline{a}(\alpha)$, if any. This upward deviation at the minimal action, or the downward deviation from $\overline{a}(\alpha)$, therefore eliminates $\beta$ as an equilibrium. Moreover, for any $\beta$ that merely selects the on-path actions in $\supp(\alpha)$, I show that if $\beta$ formed an equilibrium other than $\alpha$, then $\alpha$ would not be fully implementable. Finally, note that before the agent's incentives reach the on-path level, that is, at every $a$ with $\overline{r}(a)\precsim_{AI}r(\alpha)$, the marginal transfer satisfies $t^*_{\alpha}{'}(a)=\frac{\partial}{\partial a}u_A(a,r^*_{\alpha}(a))=\max_{a'\in[a_0,a]}\frac{\partial}{\partial a}u_A(a,r(a'))$. This means that the principal is able to price each additional unit of the agent's effort conditional on the maximal agent incentives that she has already conquered thus far. To this end, the set of offered actions $A^*_{\alpha}$ collects all those actions that induce outsider replies generating the maximum agent incentives so far, that is, every $a$ satisfying $r(a)\sim_{AI}\max_{AI}\{r(a'):a'\leq a\}$.

\begin{figure}[h]
    \centering
    \includegraphics[width=1\linewidth]{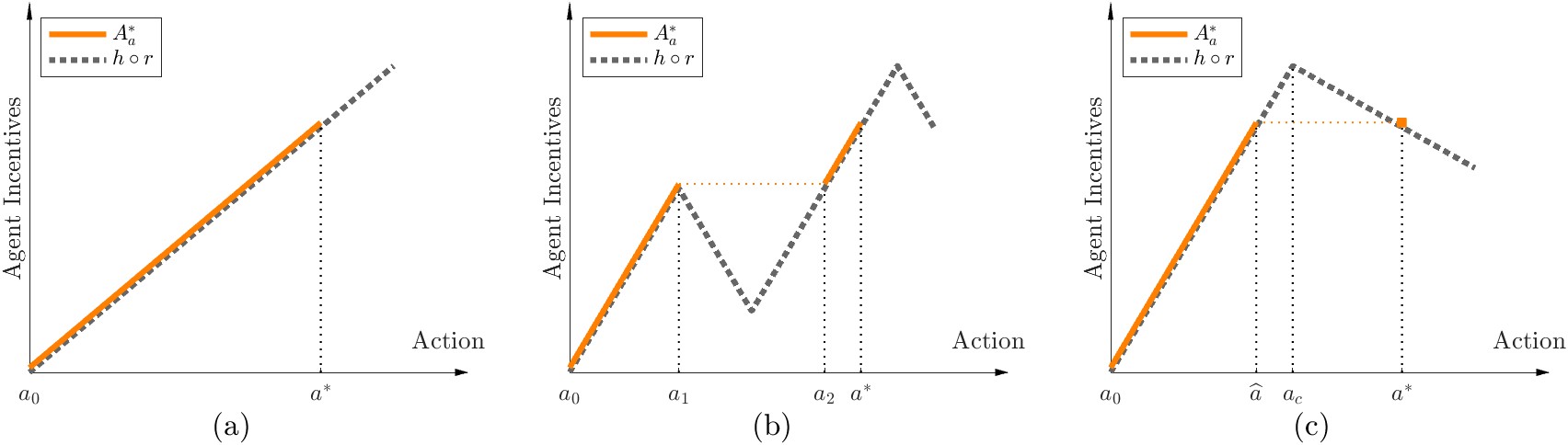}
    \caption{Robustly Optimal Contracting in Three Examples}
    \label{fig-robustly optimal contract}
\end{figure}

I illustrate the robustly optimal contract in Figure \ref{fig-robustly optimal contract} with three examples where the principal aims to implement a pure outcome $a^*$. Recall that under Assumption \ref{ranked incentives}, the order $\succsim_{AI}$ admits a continuous representation, which I denote $h$ here, that is, $r_1\succsim_{AI}r_2$ if and only if $h(r_1)\geq h(r_2)$. In Figure \ref{fig-robustly optimal contract}, $h\circ r$ maps each agent action to the agent incentive (measured in units of $h$) generated by the outsider's best reply to this action. Figure \ref{fig-robustly optimal contract}(a) depicts the classic environment of pure externalities where player actions are purely complements or substitutes. In this case, whenever the outsider expects a higher agent action, his response will move in the direction that produces higher agent incentives. Therefore, the agent's actions are \emph{self-reinforcing}, resulting in an increasing $h\circ r$. As a result, every agent action generates a greater incentive than the lower ones, so the principal offers everything in $[a_0,a^*]$ to strengthen her ability to extract surplus as quickly as possible. Figure \ref{fig-robustly optimal contract}(b) demonstrates this strategy of the principal in a more general case where agent actions can be both self-reinforcing (when $h\circ r$ increases) and self-undermining (when $h\circ r$ decreases). The principal maximally steers the agent's willingness to pay by offering only those actions that break the record of agent incentives achieved by lower actions. Consequently, she leaves a gap $[a_1,a_2]$ in the contract because offering the actions in this gap would decrease her price-raising power. Figure \ref{fig-robustly optimal contract}(c) considers a special case of such mixed externalities in which players have symmetric strategic dependence for low agent actions ($a\leq a_c$) while their strategic dependence turns asymmetric afterward ($a\geq a_c$). The application of networked competition (Section \ref{Applications}) corresponds to this structure. Here, one observes a binding equilibrium-keeping constraint (Claim \ref{claim 3}) because the principal leaves a gap $[\widehat{a},a^*]$ despite these actions generating greater agent incentives than in the desired equilibrium.

In summary, the principal uses continuous contractual terms (Figure \ref{fig-robustly optimal contract}(a)) to gain market power while robustly coordinating the action of the non-contractible party, yet she sometimes leaves gaps to avoid either a decline in market power (Figure \ref{fig-robustly optimal contract}(b)) or choking the desired outcome with excessive agent incentives (Figure \ref{fig-robustly optimal contract}(c)).

An immediate corollary is that partial implementation (Footnote \ref{partial implementation}) suffers from strategic uncertainty if and only if, compared to the outside option $a_0$, the target actions in $\alpha$ manifest self-reinforcement or, equivalently, players face \emph{overall} symmetric strategic dependence

\begin{corollary}\label{corollary-criterion}
    The partial-implementation solution is robustly optimal for inducing $\alpha\in\mathcal{A}_{FI}$ if and only if $r(\alpha)\precsim_{AI}r(a_0)$.
\end{corollary}

An example where the partial-implementation solution is robustly optimal is given by the application of regulation with boycott (Section \ref{Applications}), which manifests completely asymmetric strategic dependence, leading to agent decisions being self-undermining. To further compare the result with benchmark setups, I highlight the case of strategic substitutes as in the application of spatial competition (Section \ref{Applications}).

\paragraph{Spatial competition (revisited)}
The agent's outside option is represented by the Nash level $a_0=\frac{1}{3}$ and I consider a target action $a>a_0$. The robustly optimal contract prices effort conditional on the cumulative maximal incentives, so the (on-path) transfer is $t^*(a)=\int_{a_0}^a\left(\frac{1}{2}-\frac{3}{2}a'\right)da'$. Next, I assume that the principal partially implements the same target in the main setting, with the pricing based on the on-path outsider reaction. The transfer here is $t_P(a)=\int_{a_0}^a\left(\frac{1}{2}+\frac{a}{2}-2a'\right)da'$. One can see that $t^*(a_0)=t_P(a_0)=0$ and the marginal transfers satisfy $t^*{'}(a')<t_P'(a')$ for all $a'\in(a_0,a)$. Therefore, the coordination motive requires the principal to offer additional rents to the agent for giving up his gain in low-action equilibria.

The second benchmark with which I compare the framework is the case in which the outsider's decisions are also (bilaterally) contractible. This corresponds to the application in which a government can commit to both an industrial policy and a tariff policy, or to regulating two domestic firms insulated from international competition. The analysis follows \cite{sega2003}, who shows that with complete bilateral contractibility, strategic substitutes never cause strategic uncertainty. As a result, optimal industrial and tariff policies will each contain a single plan, yielding no gap between full and partial implementation. In sharp contrast, my result shows that losing the ability to contract with an outsider leads to strategic uncertainty despite decreasing externalities. The robustly optimal contract of Theorem \ref{main result} tackles such risks by offering a continuum of plans. This comparison therefore indicates that the regulation of firm competition tends to contain more complex contractual details when the regulator cannot contract with the outsider. This tendency is consistent with policy practice, in which industrial policies are often more involved than tariff policies (Footnote \ref{industry and tariff}), given that tariffs are less flexible or more difficult to put into effect than industrial regulatory terms (Footnote \ref{motivating spatial competition}). In other words, when governments can effectively adopt tariff policies, Segal's result implies that both policies will be simple, whereas when this is not the case and industrial policies are applied alone, Theorem \ref{main result} suggests that complex policy terms are likely to arise.

Another practical implication of my result is that the optimal design of industrial policies should be closely related to the demand structure. In both benchmarks above, the optimal contract contains a single plan, incorporating merely the demand elasticity given the outsider's on-path reply (i.e., $\frac{\partial}{\partial a}u_A(\cdot,r(a))$). However, with both non-contractibility and robustness, as in my setup, the elasticity information in every market along the outsider's best replies from the Nash level to the target level is relevant for contract design.\footnote{This point also applies to union contracting (Section \ref{Applications}). There, the robustly optimal contract takes in the information about how elastic the fringe labor supply is compared to the firm's marginal revenue.} \\

Finally, I address the implementability issue for mixed outcomes.

\begin{proposition}\label{proposition-implementability}
    For all $\alpha\in\Delta(A)$, $\alpha\in\mathcal{A}_{FI}$ if and only if $|\supp(\alpha)|\leq2$, $r(\underline{a}(\alpha))\succsim_{AI}r(\alpha)\succsim_{AI}r(\overline{a}(\alpha))$, and $\{\alpha'\in\Delta(\supp(\alpha)):r(\alpha')\sim_{AI}r(\alpha)\}=\{\alpha\}$.
\end{proposition}

Proposition \ref{proposition-implementability} shows that a fully implementable outcome can randomize over at most two agent actions. Moreover, it implies that no completely mixed outcome (with $\underline{a}(\alpha)<\overline{a}(\alpha)$) can be fully implemented in the case of pure externalities. This is because pure externalities imply $r(\overline{a}(\alpha))\succsim_{AI}r(\underline{a}(\alpha))$, so supposing $\alpha$ is fully implementable, Proposition \ref{proposition-implementability} requires $r(\underline{a}(\alpha))\sim_{AI}r(\alpha)\sim_{AI}r(\overline{a}(\alpha))$, which, however, violates the last condition in the proposition that $r(\alpha')\sim_{AI}r(\alpha)$ must have a unique solution.

\section{Optimal Outcome}\label{Robustly Optimal Outcome}

This section builds on the first-step characterization and investigates the second-step problem where the principal decides which outcome to induce. For tractability, I restrict her choice to pure outcomes, whether she implements them fully or partially. Proposition \ref{proposition-implementability} has shown that this must be true under full implementation with pure externalities because completely mixed outcomes are not fully implementable there. The main challenge of studying mixed outcomes is that, as implied by Proposition \ref{proposition-implementability}, not every outcome is implementable, adding to the principal's problem additional constraints that are hard to handle. Thus, an outcome is $(a,r(a))$, sometimes represented by $a$ alone if there is no confusion.

To uncover the effect of the principal's robustness concern, I determine how the optimal outcome under full implementation is different from that under partial implementation. Note that Proposition \ref{proposition-upper bound}, Theorem \ref{main result}, and Footnote \ref{partial implementation} have provided the principal's values $U_P^*(a)$ and $U_P^0(a)$ from fully and partially inducing each outcome $a\in A$, respectively
\begin{equation}\label{principal value}
    \begin{aligned}
        U_P^*(a)&=U_0(a)+\int_{a_0}^a\frac{\partial}{\partial a}u_A(a',r^*_a(a'))da'; \\
        U_P^0(a)&:=U_0(a)+\int_{a_0}^a\frac{\partial}{\partial a}u_A(a',r(a))da'.
    \end{aligned}
\end{equation}
By the differentiability assumption, both functions are continuous, so their sets of maximizers, $A_F:=\argmax_{a\in A}U_P^*(a)$ and $A_P:=\argmax_{a\in A}U_P^0(a)$, are nonempty.
To avoid unnecessary discussion, I further assume that strategic uncertainty is always a problem when inducing these outcomes, that is, for every $a_F\in A_F$, it holds that $r(a_F)\succ_{AI}r(a_0)$, which is the condition that Corollary \ref{corollary-criterion} identifies.\footnote{Relaxing this condition simply adds to Proposition \ref{proposition-incentive attenuation} a further case: for $a_F\in A_F$ with $r(a_F)\precsim_{AI}r(a_0)$, either $a_F\in A_P$ or $r(a_F)\precsim_{AI}r(a_P)$ for all $a_P\in A_P$. This is a harmless change, but makes the result a little harder to comprehend.}

Recall that full implementation requires the principal to give up additional rents to the agent because the transfers must be designed conditional on the agent incentives she has guaranteed thus far, instead of the on-path willingness to pay of the agent as in partial implementation. In other words, it is costly to robustly steer the agent's expectation about the outsider's reply in the principal's favor. Therefore, one might conjecture that the optimal full-implementation outcome must induce outsider replies generating no higher agent incentives than the partial solution, in order to mitigate such costs; that is, the principal may find it optimal to induce outcomes with agent incentives attenuated vis-{\`a}-vis what would be induced had she ignored strategic uncertainty. In fact, this conjecture is true regardless of the principal's utility function and the externality structure, as stated by the following result

\begin{proposition}\label{proposition-incentive attenuation}
    For all $a_F\in A_F$ and $a_P\in A_P$, $r(a_F)\precsim_{AI}r(a_P)$.
\end{proposition}

Proposition \ref{proposition-incentive attenuation} establishes that strategic rents prevent the principal from alleviating the agent's conservative skepticism about outsider behavior that results in low willingness to pay, as effectively as she can when she is able to directly select the agent's expectation. Such costs also reflect that the principal fails to internalize the agent's gain from his interaction with the outsider, contributing to an attenuation effect on agent incentives: $r(a_0)\prec_{AI}r(a_F)\precsim_{AI}r(a_P)$.

Note that this result merely speaks to the induced agent incentives in the optimal outcomes, while the comparison of agent and outsider actions is unclear. In fact, this comparison depends on the externality structure. For example, with pure externalities in Figure \ref{fig-robustly optimal contract}(a), attenuation implies $h\circ r(a_F)\leq h\circ r(a_P)$, so the strict monotonicity of $h\circ r$ further gives $a_F\leq a_P$. Thus, incentive attenuation coincides with action attenuation in this case. However, with mixed externalities, more possibilities arise, as Figure \ref{fig-robustly optimal contract}(c) illustrates. If partial implementation optimally induces the peak $a_c$, then Proposition \ref{proposition-incentive attenuation} allows a robustly optimal outcome to manifest either upward bias (such as $a^*$) or downward bias (such as $\widehat{a}$). Which bias occurs depends on the principal's objective. In the application of networked competition (Section \ref{Applications}), suppose that a platform strongly values high-quality content and rewards such content so much that low-quality content starts to crowd out high-quality creativity, which means $a_P$ lies in the region where strategic dependence is asymmetric and $h\circ r$ decreases. Then, due to robustness concerns, the platform may make the creative environment even more crowded to reduce strategic rents while obtaining excessive high-quality production.

Moreover, one might be concerned about Proposition \ref{proposition-incentive attenuation} because it does not say when the attenuation is strict (i.e., $r(a_F)\prec_{AI}r(a_P)$) and by how much the two cases differ. Answering these questions is not easy in the general case because pinning down the optimal outcomes requires knowledge of principal utility and externality structure. To provide a more precise characterization of the influence of robustness, I compare the two situations when the agent and outsider face pure externalities.

To achieve this, I introduce an \emph{integrated game} played by the principal and the outsider: imagine that the principal is choosing the action in place of the agent but with a new payoff function $\widetilde{u}_A(a,r):=u_P(a,r(a))+u_A(a,r)-u_A(a_0,r)$, while the outsider still picks his decision with utility $u_O$. Two equilibrium concepts are considered for different timings of the integrated game. An outcome is \emph{Stackelberg} if it is the on-path play of an SPNE of the game with the principal moving first and the outsider following. Moreover, an outcome is \emph{Nash} if it is played in a Nash equilibrium of the game with both players moving simultaneously. A \emph{principal-preferred Nash} outcome maximizes the principal's maximal payoff guarantee $U_P^*(\cdot)$ among all Nash outcomes. Note that, by definition, all Stackelberg outcomes give the principal the same payoff in the integrated game. The following result matches the partial and full solutions to these concepts

\begin{proposition}\label{proposition-integrated game}
    Every $a_P\in A_P$ is a Stackelberg outcome of the integrated game. With pure externalities, every $a_F\in A_F$ is a principal-preferred Nash outcome of the integrated game.
\end{proposition}

The former part of Proposition \ref{proposition-integrated game} is straightforward, since a partially implementing principal selects $a$ to maximize $\widetilde{u}_A(a,r(a))$, accounting for the outsider's best response. This is consistent with the setting in which the contract is public, so, absent an equilibrium selection issue, this commitment power enables the principal to be a leader. To see the latter part, note the following: for all $a\in A$,
\begin{equation*}
    \begin{aligned}
        \widetilde{u}_A(a_F,r(a_F&))-\widetilde{u}_A(a,r(a_F))=u_P(a_F,r(a_F))-u_P(a,r(a))+u_A(a_F,r(a_F))-u_A(a,r(a_F)) \\
        &=\left[U_P^*(a_F)-U_P^*(a)\right]-\int_a^{a_F}\frac{\partial}{\partial a}u_A(a',r(a'))da'+\int_a^{a_F}\frac{\partial}{\partial a}u_A(a',r(a_F))da' \\
        &\geq\int_a^{a_F}\left[\frac{\partial}{\partial a}u_A(a',r(a_F))-\frac{\partial}{\partial a}u_A(a',r(a'))\right]da'\geq0,
    \end{aligned}
\end{equation*}
where the second equality applies the definition (\ref{principal value}) and the fact that pure externalities imply $r^*_{\widehat{a}}(a')=r(a')$ for all $\widehat{a}$ and $a'\leq\widehat{a}$; the first inequality results from $a_F$ maximizing $U_P^*(\cdot)$; the last inequality holds because $r(a')\succsim_{AI}r(a_F)$ if $a_F<a$ while $r(a')\precsim_{AI}r(a_F)$ if $a_F>a$, as a result of pure externalities. Consequently, $a_F$ is a best response to $r(a_F)$, which in turn optimally replies to $a_F$, forming a Nash equilibrium. Furthermore, $a_F$ must also be a principal-preferred Nash outcome because it maximizes $U_P^*(\cdot)$ among all outcomes and thus among all Nash outcomes.

Proposition \ref{proposition-integrated game} reveals a sharp comparison between full and partial implementation. On the one hand, the robustness concern costs the principal her entire commitment value in the integrated game, since she induces outcomes as if she did not lead the outsider. On the other hand, the ability to commit to a public contract enables the principal to select her favorite outcome among the Nash ones. Therefore, in the integrated game, robustness turns the principal's commitment power into the power to select among Nash outcomes, which coordination power alone would give her; in this sense, commitment power and coordination power are substitutes in terms of influencing economic output. When the principal's payoff $u_P$ depends only on the agent's action, the integrated game also describes private contracting, as explained next: a principal who can commit yet lacks the ability to coordinate equilibrium play in her favor then chooses among the same Nash outcomes as a principal who can select the equilibrium while being unable to commit, and the two powers are perfect substitutes when the Nash outcome is unique. With several Nash outcomes, their choices may still differ, since only the one who can commit pays strategic rents.

In fact, a principal lacking commitment power can be viewed as offering a contract that will not be observed by the outsider. In this case, the principal designs the contract conditional on a fixed outsider decision that she expects in equilibrium. As a result, when her payoff $u_P$ depends only on the agent's action, the optimal outcomes there are exactly all the Nash outcomes of the integrated game, whether implementation is full or partial. This setting is referred to as \emph{private contracting}, while the main model assumes that the contract is publicly observable to the outsider. Surprisingly, the principal could be strictly worse off with public contracting than with private contracting: in the main model, the principal often pays positive strategic rents and earns the robust value $U_P^*(\cdot)$, whereas she captures the partial value $U_P^0(\cdot)$ when the contract is hidden from the outsider. For example, with a unique Nash outcome and a payoff $u_P$ that depends only on the agent's action, Proposition \ref{proposition-integrated game} implies that private contracting yields strictly higher principal surplus than public contracting under full implementation. In sharp contrast, public contracting under partial implementation is never worse for the principal than private contracting because a principal in the former setting gains her commitment value by inducing a Stackelberg outcome. Therefore, if contractual transparency is within the principal's discretion, accounting for strategic robustness and ignoring it can generate opposite predictions about which institution would endogenously emerge.

Finally, combining the forces I identify in this and the previous sections, I conclude that addressing strategic uncertainty dampens principal welfare through two channels: she pays strategic rents and, with pure externalities, loses her entire commitment value in the integrated game.

\section{Discussion}\label{Discussions}

\paragraph{Full access} The main setup assumes that without a contract, the agent can merely access his outside option. This may be less reasonable in certain scenarios, such as spatial competition (Section \ref{Applications}), where a domestic firm has full discretion over its production. To give the agent full access to his actions, I further require that every contract $M$ proposed by the principal contain all plans $(a,0)\in M$ with $a\in A:=[\underline{a},\overline{a}]$. Moreover, I assume that when the principal is absent, the agent and outsider will play a unique pure-strategy Nash equilibrium (e.g., a unique Cournot outcome in spatial competition without regulation). This ``natural state'', denoted as $(a_0,r(a_0))$, plays the same role as the outside option in the main setting. In addition, the agent's action $a^*$ guaranteed by the principal can now be on either side of $a_0$, that is, one can have either $a^*>a_0$ or $a^*<a_0$.

In this extension, I argue that the first-step characterization, Theorem \ref{main result}, still holds when the externalities are pure and the agent's payoff is weakly concave in his action. I explain this by discussing the following three issues. First, when the principal guarantees an action lower than $a_0$, the analysis is identical to that of inducing an action higher than $-a_0$ after relabeling the agent's action $a$ as $-a$. For Theorem \ref{main result} to apply, the main issue is whether Assumption \ref{ranked incentives} and Assumption \ref{intuitive mixture} will continue to hold. The answer is affirmative: (i) pure externalities are still pure after adding a negative sign to $a$; (ii) the new order of agent incentives $\succsim_{-AI}$ defined for $-a$ is still quasi-concave, since pure externalities imply that the original order $\succsim_{AI}$ is monotonic and thus quasi-convex. Second, the proof of Claim \ref{claim 4} suggests that, in this case, the upper bound for each $R(a;M)$ should instead be the maximal agent incentives rationalized by actions in $[\underline{a},a]$, rather than $[a_0,a]$. However, with pure externalities, this is not a problem because $r(a)\succsim_{AI}r(a')$ for all $a'\leq a$. Lastly, the agent's full access to actions imposes an additional constraint: no positive transfer is effective since the agent can simply reject the contract and select the same action without being punished by such a transfer. This is not a problem if Theorem \ref{main result} gives non-positive transfers, as in the example that I considered in Section \ref{Main Result} when revisiting the application of spatial competition. Under the above conditions, one can show that Theorem \ref{main result} always gives non-positive transfers and that the added plans $(a,0)$ do not affect full implementation. Beyond these conditions, one would need to add the constraint $t\leq0$, along with the two upper bounds in Claim \ref{claim 3} and Claim \ref{claim 4}, suitably modified, to the principal's problem. I conjecture that the construction of the optimal transfer function (\ref{optimal contract}) is then modified as follows
\begin{equation*}
    t^*_{\alpha}{'}(a):=\begin{cases}
        \min\left\{\frac{\partial}{\partial a}u_A(a,r^*_{\alpha}(a)),0\right\} & \text{, if }t^*_{\alpha}(a)=0; \\
        \frac{\partial}{\partial a}u_A(a,r^*_{\alpha}(a)) &\text{, if }t^*_{\alpha}(a)<0.
    \end{cases}
\end{equation*}
The above modification demonstrates the principal's intention to raise agent incentives and thus transfers as much as possible subject to the three constraints. Consequently, the duality approach would extend to this full-access case in general, with the main result modified as above.

\bibliographystyle{aer}
\bibliography{bib}

\appendix
\renewcommand{\thesection}{Appendix \Alph{section}}
\renewcommand{\thesubsection}{\Alph{section}.\arabic{subsection}}

\section{}\label{Appendix A}

Recall that Assumption \ref{ranked incentives} admits a continuous representation of $\succsim_{AI}$, which I denote as $h$, such that $r_1\succsim_{AI}r_2$ if and only if $h(r_1)\geq h(r_2)$. I maintain this notation in all proof sections.

I also record the following lemma, which shows that outsider actions equivalent in the order of agent incentives are interchangeable in both problems (\ref{V(r;M)}) and (\ref{T(a;M)}).

\begin{lemma}\label{lemma-class}
    For all $r_1,r_2\in R$ with $r_1\sim_{AI}r_2$, $u_A(a,r_1)-u_A(a,r_2)$ does not depend on $a\in A$. Hence, for every $M\in\mathcal{M}$, a plan in $M$ is optimal given $r_1$ if and only if it is optimal given $r_2$, $V(r_1;M)-V(r_2;M)=u_A(a,r_1)-u_A(a,r_2)$ for all $a\in A$, and the objective function of problem (\ref{T(a;M)}) takes the same value at $r_1$ and $r_2$ for every $a\in A$. In particular, $r_1\in R(a;M)$ if and only if $r_2\in R(a;M)$, so each $R(a;M)$ is a union of equivalence classes of $\sim_{AI}$.
\end{lemma}

To see this, note that $r_1\sim_{AI}r_2$ means both $r_1\succsim_{AI}r_2$ and $r_2\succsim_{AI}r_1$, that is, $\frac{\partial u_A}{\partial a}(a,r_1)\geq\frac{\partial u_A}{\partial a}(a,r_2)$ and $\frac{\partial u_A}{\partial a}(a,r_1)\leq\frac{\partial u_A}{\partial a}(a,r_2)$ for all $a$, so the two derivatives coincide. Integrating from $a_0$ yields $u_A(a,r_1)-u_A(a,r_2)=u_A(a_0,r_1)-u_A(a_0,r_2)=:c$ for all $a\in A$. Thus, $u_A(a',r_1)-t'=u_A(a',r_2)-t'+c$ for every $(a',t')\in M$, so the objective functions of problem (\ref{V(r;M)}) given $r_1$ and $r_2$ differ by the constant $c$, which yields the same optimal plans and $V(r_1;M)=V(r_2;M)+c$. Consequently, $u_A(a,r_1)-V(r_1;M)=u_A(a,r_2)-V(r_2;M)$ for all $a\in A$, which establishes the remaining statements.

\subsection{Proof of Claim \ref{claim 1}}

In fact, I show the stated value is earned by the principal in any equilibrium induced by any contract, say $M$. It suffices to show that every plan $(a^*,t^*)$ chosen on the equilibrium path satisfies $t^*=T(a^*;M)$. Note that the optimality of $(a^*,t^*)$ given the outsider decision $r(\alpha)$ requires
\[V(r(\alpha);M)=\max_{(a',t')\in M}u_A(a',r(\alpha))-t'=u_A(a^*,r(\alpha))-t^*,\]
which further implies the following lower bound for $T(a^*;M)$
\[T(a^*;M)=\max_{r'}u_A(a^*,r')-V(r';M)\geq u_A(a^*,r(\alpha))-V(r(\alpha);M)=t^*.\]
On the other hand, by replacing $M$ with a singleton subset $\{(a^*,t^*)\}$, one has
\[V(r;M)=\max_{(a',t')\in M}u_A(a',r)-t'\geq\max_{(a',t')\in\{(a^*,t^*)\}}u_A(a',r)-t'=u_A(a^*,r)-t^*,\]
from which I show that the lower bound above is also an upper bound for $T(a^*;M)$
\begin{equation}\label{T>t}
    T(a^*;M)=\max_{r'}u_A(a^*,r')-V(r';M)\leq\max_{r'}u_A(a^*,r')-u_A(a^*,r')+t^*=t^*.
\end{equation}
This completes the proof of Claim \ref{claim 1} because, from the above, I conclude $T(a^*;M)=t^*$. More generally, the above proof also establishes the following lemma
\begin{lemma}\label{lemma-r in R}
    For all $M\in\mathcal{M}$ and $r\in R$, if $V(r;M)$ is achieved at $(a,t)$, then $r\in R(a;M)$ and $T(a;M)=t$.
\end{lemma}

\subsection{Proof of Claim \ref{claim 2}}

Fix any $M\in\mathcal{M}$. The proof proceeds by showing that $T(\cdot;M)$ is absolutely continuous and that the derivative form of the envelope theorem holds almost everywhere. To begin with, note that the twice-differentiability of $u_A$ on a compact set $A\times R$ implies that it is Lipschitz in both $a$ and $r$, with Lipschitz constants $L_a$ and $L_r$, respectively. I first show $V(\cdot;M)$ is Lipschitz with constant $L_r$: for all $r,r'\in R$ and $(a,t)\in M$,
\begin{align*}
    &u_A(a,r)-t\leq u_A(a,r')-t+L_r|r-r'|\leq V(r';M)+L_r|r-r'| \\
    \Rightarrow\;&V(r;M)=\max_{(a,t)\in M}u_A(a,r)-t\leq V(r';M)+L_r|r-r'| \\
    \Rightarrow\;&\text{by symmetry, also }V(r';M)\leq V(r;M)+L_r|r-r'|.
\end{align*}
Similarly, $T(\cdot;M)$ is Lipschitz with constant $L_a$: for all $a,a'\in A$ and $r\in R$,
\begin{align*}
    &u_A(a,r)-V(r;M)\leq u_A(a',r)-V(r;M)+L_a|a-a'|\leq T(a';M)+L_a|a-a'| \\
    \Rightarrow\;&T(a;M)=\max_{r\in R}u_A(a,r)-V(r;M)\leq T(a';M)+L_a|a-a'| \\
    \Rightarrow\;&\text{by symmetry, also }T(a';M)\leq T(a;M)+L_a|a-a'|.
\end{align*}
Hence, the absolute continuity of $T(\cdot;M)$ is implied by its Lipschitz continuity on the interval $A$.

Second, I show that the derivative form of the envelope theorem holds almost everywhere. Since $T(\cdot;M)$ is Lipschitz continuous, Rademacher's theorem ensures that it is differentiable almost everywhere on $A$. Take any interior $a$ at which $T(\cdot;M)$ is differentiable and any $\widetilde{r}(a)\in R(a;M)$, where $R(a;M)$ is nonempty because $u_A(a,\cdot)-V(\cdot;M)$ is continuous on the compact set $R$. By the definition (\ref{T(a;M)}), $T(a';M)\geq u_A(a',\widetilde{r}(a))-V(\widetilde{r}(a);M)$ for all $a'\in A$, with equality at $a'=a$. That is, the function $a'\mapsto T(a';M)-u_A(a',\widetilde{r}(a))+V(\widetilde{r}(a);M)$ is nonnegative and attains its minimum value zero at the interior point $a$. Since this function is differentiable at $a$, its derivative vanishes there, which gives $\frac{\partial}{\partial a}T(a;M)=\frac{\partial}{\partial a}u_A(a,\widetilde{r}(a))$. As this holds for every $\widetilde{r}(a)\in R(a;M)$, for any selection $\widetilde{r}(a')\in R(a';M)$, the integrand $\frac{\partial}{\partial a}u_A(a',\widetilde{r}(a'))$ coincides with the derivative of $T(\cdot;M)$ almost everywhere, so it is measurable and bounded. Therefore, the absolute continuity of $T(\cdot;M)$ yields $T(a;M)=T(a_0;M)+\int_{a_0}^a\frac{\partial}{\partial a}u_A(a',\widetilde{r}(a'))da'$ for all $a\in A$, which completes the proof.

\subsection{Proof of Claim \ref{claim 3}}

As already explained in the main text, it suffices to show $R(\cdot;M)$ is nondecreasing in the order of agent incentives. That is, I show the following lemma

\begin{lemma}\label{lemma-mono R}
    For all $M\in\mathcal{M}$ and $a_1,a_2\in A$ with $a_1<a_2$, $R(a_1;M)\precsim_{AI}R(a_2;M)$.
\end{lemma}

To prove this, I take any $a_1,a_2\in A$ with $a_1<a_2$, $r_1\in R(a_1;M)$, and $r_2\in R(a_2;M)$. For the sake of contradiction, suppose that $r_1\succ_{AI}r_2$. Optimality in problem (\ref{T(a;M)}) requires
\begin{align*}
    u_A(a_1,r_1)-V(r_1;M)\geq u_A(a_1,r_2)-V(r_2;M); \\
    u_A(a_2,r_2)-V(r_2;M)\geq u_A(a_2,r_1)-V(r_1;M).
\end{align*}
Summing up both sides and canceling the repeated terms yield
\[\int_{a_1}^{a_2}\left[\frac{\partial}{\partial a}u_A(a',r_1)-\frac{\partial}{\partial a}u_A(a',r_2)\right]da'\leq0,\]
which contradicts $a_2>a_1$ and $r_1\succ_{AI}r_2$, since the latter makes the integrand strictly positive by Assumption \ref{ranked incentives}. Therefore, one must have $r_1\precsim_{AI}r_2$ and thus $R(a_1;M)\precsim_{AI}R(a_2;M)$. Finally, Claim \ref{claim 3} follows. Fix any $M$ that fully implements $\alpha$. Since $\supp(\beta_M)$ is compact, its projection onto actions is closed and has full $\alpha$-measure, so it contains $\supp(\alpha)$ and, in particular, $\overline{a}(\alpha)$; that is, some plan $(\overline{a}(\alpha),t)$ is chosen on the equilibrium path. This plan is optimal given the outsider's equilibrium decision $r(\alpha)$, so Lemma \ref{lemma-r in R} gives $r(\alpha)\in R(\overline{a}(\alpha);M)$, and Lemma \ref{lemma-mono R} then gives $R(a;M)\precsim_{AI}r(\alpha)$ for every $a<\overline{a}(\alpha)$.

\subsection{Proof of Claim \ref{claim 4}}

To start with, I show the following four lemmas

\begin{lemma}\label{lemma-two sides}
    For all $M\in\mathcal{M}$ and $a\in A$, if $r\in R(a;M)$ and there is $r'$ such that $r\succ_{AI}[\prec_{AI}]r'$, $V(r;M)$ can be achieved at some $(\widetilde{a},\widetilde{t})$ with $\widetilde{a}\leq[\geq]a$.
\end{lemma}

To establish Lemma \ref{lemma-two sides}, I begin with the case where $r\succ_{AI}r'$, while the other case, where $r\prec_{AI}r'$, follows from symmetric arguments. For the sake of contradiction, suppose that there is no such $\widetilde{a}$ with $\widetilde{a}\leq a$, meaning that the agent picks only actions larger than $a$ given $r$. Since the objective function of problem (\ref{V(r;M)}) is jointly continuous in the plan and the outsider action, and $M$ is compact, Berge's maximum theorem implies that, given each outsider action, the set of optimal plans is nonempty and compact, and that this set is upper-hemicontinuous in the outsider action. Next, since $r\succ_{AI}r'$ and $h$ is continuous on the interval $R$, the intermediate value theorem yields a sequence $(r_n)_{n=1}^{\infty}\subset R$ with $h(r')<h(r_n)<h(r)$ for all $n$ and $h(r_n)\rightarrow h(r)$, so each $r_n\prec_{AI}r$. As $R$ is compact, a subsequence, still denoted $(r_n)_{n=1}^{\infty}$, converges to some $\widetilde{r}\in R$, and the continuity of $h$ gives $h(\widetilde{r})=h(r)$, that is, $\widetilde{r}\sim_{AI}r$. By Lemma \ref{lemma-class}, the plans optimal given $\widetilde{r}$ are exactly those optimal given $r$, which all lie in the set $\{(a',t')\in M:a'>a\}$ by supposition. Since this set is open relative to $M$, upper-hemicontinuity at $\widetilde{r}$ implies that, for a sufficiently large $n$, every plan optimal given $r_n$ also has an action larger than $a$. I fix such an $n$ and let $r'':=r_n$, so $r''\prec_{AI}r$ and every plan $(\widehat{a}'',\widehat{t}'')$ optimal given $r''$ satisfies $\widehat{a}''>a$. I also take any plan $(\widehat{a},\widehat{t})$ optimal given $r$. The following then holds
\begin{align*}
    u_A(a,r'')-V(r'';M)&=u_A(a,r'')-u_A(\widehat{a}'',r'')+\widehat{t}''>u_A(a,r)-u_A(\widehat{a}'',r)+\widehat{t}'' \\
    &\geq u_A(a,r)-u_A(\widehat{a},r)+\widehat{t}=u_A(a,r)-V(r;M),
\end{align*}
where the first inequality utilizes $\widehat{a}''>a$ and $r''\prec_{AI}r$, and the second inequality results from the optimality of $(\widehat{a},\widehat{t})$ given $r$. One can see that this forms a contradiction to $r\in R(a;M)$. Therefore, one must have such an $\widetilde{a}$ with $\widetilde{a}\leq a$.

\begin{lemma}\label{lemma-convexity}
    For all $M\in\mathcal{M}$, $a\in A$, and $r_1,r_2\in R$ with $r_1\prec_{AI}r_2$, if $r_1,r_2\in R(a;M)$, any $r$ such that $r_1\prec_{AI}r\prec_{AI}r_2$ is also in $R(a;M)$.
\end{lemma}

To prove Lemma \ref{lemma-convexity}, I take $r_1$, $r_2$, and $r$ as stated. Lemma \ref{lemma-two sides} then implies that there exists a plan $(a_1,t_1)$ that is optimal given $r_1$ and satisfies $a_1\geq a$. Likewise, there is a plan $(a_2,t_2)$ that is optimal given $r_2$ and satisfies $a_2\leq a$. However, one must have $a_1=a_2=a$ because, otherwise, $a_1>a_2$, along with $r_2\succ_{AI}r_1$, would contradict the incentive compatibility conditions
\begin{align*}
    &u_A(a_1,r_1)-t_1\geq u_A(a_2,r_1)-t_2;\qquad u_A(a_2,r_2)-t_2\geq u_A(a_1,r_2)-t_1 \\
    \Rightarrow\;&\text{summing up both sides yields }\int_{a_1}^{a_2}\left[\frac{\partial}{\partial a}u_A(a',r_2)-\frac{\partial}{\partial a}u_A(a',r_1)\right]da'\geq0.
\end{align*}
Therefore, $V(r_1;M)$ and $V(r_2;M)$ are both achieved at a critical plan $(a,t)$ (defined below), where $t:=t_1=t_2$ because Lemma \ref{lemma-r in R} gives $t_1=T(a;M)=t_2$, so $T(a;M)=t$. Note that $(a,t)$ is also optimal given $r$ because all plans with actions larger than $a$ are weakly less attractive relative to $(a,t)$ under $r$ than under $r_2$, while those with actions smaller than $a$ are weakly less attractive relative to $(a,t)$ under $r$ than under $r_1$, so $(a,t)$ being critical and optimal under $r_1$ and $r_2$ implies its optimality under $r$. As a result, the principal finds it optimal to pick $r$ in problem (\ref{T(a;M)}) since it induces the agent to choose $(a,t)$ and thus $u_A(a,r)-V(r;M)=t=T(a;M)$, attaining the optimal transfer.

A useful remark is that Lemma \ref{lemma-convexity} reveals the convexity of each $R(a;M)$, so, by Lemma \ref{lemma-class} (an outsider action equivalent to an element of $R(a;M)$ also belongs to $R(a;M)$), the relationship of $R(a;M)$ with any $r\in R$ must fall into one of three cases: either $r\succ_{AI}R(a;M)$, $r\prec_{AI}R(a;M)$, or $r\in R(a;M)$.

\begin{lemma}\label{lemma-equilibrium}
    For all $M\in\mathcal{M}$ and $(a,t)\in M$ that achieves $V(r;M)$ for some $r\in R$, if $r(a)\in R(a;M)$, $(a,t)$ forms an equilibrium.
\end{lemma}

In all proofs that follow, I call a plan $(a,t)$ that is optimal given some $r$, as in Lemma \ref{lemma-equilibrium}, a \emph{critical plan}.

To demonstrate Lemma \ref{lemma-equilibrium}, I take any critical plan $(a,t)$ and let $R_0$ collect all outsider actions under which $(a,t)$ is optimal. The goal is thus to show $r(a)\in R_0$. To begin with, I show $R_0$ is convex, that is, whenever $r_1,r_2\in R_0$ and $r_1\precsim_{AI}r_2$, any $r$ with $r_1\precsim_{AI}r\precsim_{AI}r_2$ is also in $R_0$. This is because all plans with actions larger than $a$ are weakly less attractive relative to $(a,t)$ under $r$ than under $r_2$, while those with actions smaller than $a$ are weakly less attractive relative to $(a,t)$ under $r$ than under $r_1$, so $(a,t)$ being critical and optimal under $r_1$ and $r_2$ implies its optimality under $r$. Next, suppose $r(a)\succ_{AI}R_0$. Notice that the case with $r(a)\prec_{AI}R_0$ is symmetric. After I rule out these two cases, the convexity of $R_0$ must yield $r(a)\in R_0$. Let $(a',t')$ be a plan that is optimal given $r(a)$. One must have $a'\geq a$ because, for any $r'\in R_0$, the incentive compatibility conditions require
\begin{align*}
    &u_A(a,r')-t\geq u_A(a',r')-t';\qquad u_A(a',r(a))-t'\geq u_A(a,r(a))-t \\
    \Rightarrow\;&\text{summing up both sides yields }\int_{a}^{a'}\left[\frac{\partial}{\partial a}u_A(s,r(a))-\frac{\partial}{\partial a}u_A(s,r')\right]ds\geq0,
\end{align*}
which shows that $a'<a$ would contradict $r(a)\succ_{AI}r'$. Moreover, $a'\neq a$ because $(a,t)$ is critical, so $a'=a$ implies $t'=t$ (Lemma \ref{lemma-critical}), meaning that $(a,t)$ best replies to $r(a)$ and thus $r(a)\in R_0$, a contradiction. Hence, $a'>a$. This means one cannot find such an $a'$ with $a'\leq a$, which contradicts Lemma \ref{lemma-two sides} because one has $r(a)\in R(a;M)$ and $r(a)\succ_{AI}r'$. In summary, one must instead have $r(a)\in R_0$. \\

\begin{lemma}\label{lemma-critical}
    For all $M\in\mathcal{M}$, the following hold: (i) the set of critical plans in $M$ is nonempty and compact; (ii) two critical plans with the same action have the same transfer; (iii) for all $r,r'\in R$ with $r'\succ_{AI}r$, if $(a',t')\in M$ is optimal given $r'$ and $(a,t)\in M$ is optimal given $r$, then $a'\geq a$; (iv) the critical plan with the lowest [highest] action is optimal given $\min_{AI}R$ [$\max_{AI}R$].
\end{lemma}

To prove Lemma \ref{lemma-critical}, I fix $M\in\mathcal{M}$. For (i), since $M$ is nonempty and compact and $u_A(a,r)-t$ is continuous in $(a,t,r)$, $V(r;M)$ is attained for every $r\in R$, so critical plans exist. Moreover, the graph $G:=\{(r,a,t)\in R\times M:(a,t)\text{ is optimal given }r\}$ is closed: if $(r_n,a_n,t_n)\in G$ converges to $(r,a,t)$, then $(a,t)\in M$ since $M$ is closed, and taking limits in $u_A(a_n,r_n)-t_n\geq u_A(a'',r_n)-t''$ for each $(a'',t'')\in M$ shows that $(a,t)$ is optimal given $r$ (this is the upper-hemicontinuity in Berge's maximum theorem). Hence, $G$ is a closed subset of the compact set $R\times M$, so $G$ is compact, and so is its projection onto $M$, which is exactly the set of critical plans. For (ii), if $(a,t)$ and $(a,\widehat{t})$ are both critical with $t<\widehat{t}$, then $(a,t)$ gives the agent a strictly higher payoff than $(a,\widehat{t})$ under every $r$, so $(a,\widehat{t})$ is never optimal, a contradiction. For (iii), summing up the optimality conditions $u_A(a',r')-t'\geq u_A(a,r')-t$ and $u_A(a,r)-t\geq u_A(a',r)-t'$ yields
\[\int_{a'}^{a}\left[\frac{\partial}{\partial a}u_A(s,r)-\frac{\partial}{\partial a}u_A(s,r')\right]ds\geq0,\]
whose integrand is strictly negative by Assumption \ref{ranked incentives} because $r'\succ_{AI}r$, so $a'<a$ is impossible. For (iv), the critical plan with the lowest action, denoted $(a_l,t_l)$, exists by (i) and is unique by (ii), and it is optimal given some $r$. Let $(\widehat{a},\widehat{t})$ be any plan that is optimal given $\min_{AI}R$; this plan is critical. If $r\sim_{AI}\min_{AI}R$, then $(a_l,t_l)$ is optimal given $\min_{AI}R$ by Lemma \ref{lemma-class}. Otherwise, $r\succ_{AI}\min_{AI}R$, so (iii) gives $a_l\geq\widehat{a}$, whereas $\widehat{a}\geq a_l$ by the definition of $a_l$; hence, $\widehat{a}=a_l$, and (ii) yields $(\widehat{a},\widehat{t})=(a_l,t_l)$, which is thus optimal given $\min_{AI}R$. The case of the critical plan with the highest action is symmetric.

Now, I return to proving Claim \ref{claim 4}. I fix a contract $M$ that fully implements some $\alpha$. To maintain notational conciseness, I write $a^*$ for $\underline{a}(\alpha)$ in the rest of the proof. The first step is to suppose that there is some $a\in[a_0,a^*)$ with $\overline{r}(a)\prec_{AI} R(a;M)$ and thereby construct an under-action equilibrium. The second step is to further show that even when $\overline{r}(a)\in R(a;M)$, any $r''\in R(a;M)$ with $r''\succ_{AI}\overline{r}(a)$ leads to an under-action equilibrium. These suffice due to the convexity of $R(a;M)$ given by Lemma \ref{lemma-convexity} and Lemma \ref{lemma-class}. \\

Step 1: suppose that there is some $a\in[a_0,a^*)$ with $\overline{r}(a)\prec_{AI} R(a;M)$.

By considering two cases, I find some $a'<a^*$ such that $a'$ is offered in $M$ and $r(a')\prec_{AI}R(a';M)$. To do this, let the maximal value $T(a;M)$ be attained at some $r''$, and the maximal value $V(r'';M)$ be attained at some $(a'',t'')$. First, if some such $(a'',t'')$ has $a''=a$, then $a$ is exactly this $a'$ that I intend to find since $a''$ is offered in $M$, $a<a^*$, and $r(a)\precsim_{AI}\overline{r}(a)\prec_{AI}R(a;M)$. In contrast, if no such $(a'',t'')$ has $a''=a$, then Lemma \ref{lemma-two sides} ensures that there must be such an $a''$ that is lower than $a$ because $\overline{r}(a)\prec_{AI}r''\in R(a;M)$. Moreover, Lemma \ref{lemma-r in R} indicates that $a''$ being optimal given $r''$ implies that $r''\in R(a'';M)$. Since $a''<a$ and $r''\in R(a;M)$, one has $r(a'')\precsim_{AI}\overline{r}(a)\prec_{AI}r''$. Next, suppose that $r(a'')\prec_{AI}R(a'';M)$ is not true, meaning that there is some $\widehat{r}\in R(a'';M)$ such that $\widehat{r}\precsim_{AI}r(a'')$. However, one also has $r(a'')\prec_{AI}r''$ and $r''\in R(a'';M)$, so Lemma \ref{lemma-convexity} (together with Lemma \ref{lemma-class} if $\widehat{r}\sim_{AI}r(a'')$) yields $r(a'')\in R(a'';M)$, which, by Lemma \ref{lemma-equilibrium}, implies that this critical plan offering $a''$ forms an equilibrium, contradicting full implementation. Consequently, one must have $r(a'')\prec_{AI}R(a'';M)$, and thus the desired action can be set as $a'=a''$.

For notational conciseness, I simply use $(a,t)$ in place of $(a',t')$ from now on. One can see from the above that $(a,t)$ is a critical plan offered in $M$ that satisfies $a<a^*$ and $r(a)\prec_{AI} R(a;M)$.

Next, I argue that the critical plan with the lowest action, denoted $(a_l,t_l)$ (which exists and is unique by Lemma \ref{lemma-critical}; note that $a_l\leq a<a^*$), has $r(a_l)\succ_{AI}R(a_l;M)$. This is because if $r(a_l)\in R(a_l;M)$, Lemma \ref{lemma-equilibrium} implies that this plan forms an undesired equilibrium; on the other hand, $r(a_l)\prec_{AI}R(a_l;M)$ cannot happen since this plan must best reply to the outsider action that generates the lowest agent incentives, $r_l:=\min_{AI} R$ (Lemma \ref{lemma-critical}), so Lemma \ref{lemma-r in R} gives $r_l\in R(a_l;M)$; however, it is impossible to have $r(a_l)\prec_{AI}r_l\in R(a_l;M)$.

Now, I am ready to construct an under-action equilibrium. Recall that any critical plan with action $a'<a^*$ cannot form an equilibrium, so it cannot have $r(a')\in R(a';M)$; hence, one must have either $r(a')\prec_{AI}R(a';M)$ or $r(a')\succ_{AI}R(a';M)$, as implied by Lemmas \ref{lemma-convexity}, \ref{lemma-equilibrium}, and \ref{lemma-class}. I first define the following set
\begin{equation*}
    \begin{aligned}
        \underline{M}:=\{(a',t'):&(a',t')\text{ is critical, and every critical plan} \\
        &(\widetilde{a},\widetilde{t})\text{ with }        \widetilde{a}\in[a_l,a']\text{ has }r(\widetilde{a})\succ_{AI}R(\widetilde{a};M)\}.
    \end{aligned}
\end{equation*}
In other words, $\underline{M}$ contains all critical plans that are close to $a_l$ and have the same relationship between $R(\cdot;M)$ and $r(\cdot)$ as with $a_l$. $\underline{M}$ is nonempty because $(a_l,t_l)\in\underline{M}$, and it cannot contain any plan with action $a'\geq a$ because $r(a)\prec_{AI}R(a;M)$. Let the supremum action be $a_1:=\sup\{a':a'\text{ is offered in }\underline{M}\}$. Since the set of critical plans is compact (Lemma \ref{lemma-critical}), $a_1$ is a critical action, offered with a unique transfer $t_1$ (Lemma \ref{lemma-critical}), and $a_1\leq a$. In addition, the objective of problem (\ref{T(a;M)}) is continuous in $(a,r)$ because $V(\cdot;M)$ is Lipschitz (see the proof of Claim \ref{claim 2}), so Berge's maximum theorem implies that $R(\cdot;M)$ is upper-hemicontinuous with nonempty compact values. Since $h\circ r$ is continuous, the function $a'\mapsto h(r(a'))-\max_{r'\in R(a';M)}h(r')$ is lower semicontinuous and the function $a'\mapsto h(r(a'))-\min_{r'\in R(a';M)}h(r')$ is upper semicontinuous, so the sets $A_{\succ}:=\{a'\in A:r(a')\succ_{AI}R(a';M)\}$ and $A_{\prec}:=\{a'\in A:r(a')\prec_{AI}R(a';M)\}$ are open in $A$. I discuss two cases. First of all, if $a_1$ is not offered in $\underline{M}$, then one has $r(a_1)\prec_{AI}R(a_1;M)$ because every critical action in $[a_l,a_1)$ is lower than the action of some plan in $\underline{M}$ and thus belongs to $A_{\succ}$, so $(a_1,t_1)\notin\underline{M}$ requires $a_1\notin A_{\succ}$, while the critical action $a_1\leq a<a^*$ cannot have $r(a_1)\in R(a_1;M)$; in addition, I obtain a sequence of plans $(a_n,t_n)_{n=1}^{\infty}\subset\underline{M}$ such that $a_n\rightarrow a_1$. However, this is impossible because $r(\cdot)$ is continuous, $\succsim_{AI}$ is continuous, and $R(\cdot;M)$ is upper-hemicontinuous, so $r(\cdot)\succ_{AI}R(\cdot;M)$ cannot suddenly jump to $r(\cdot)\prec_{AI}R(\cdot;M)$ at a single point; more precisely, $a_1$ belongs to the open set $A_{\prec}$, so $a_n\in A_{\prec}$ for large $n$, contradicting $a_n\in A_{\succ}$, which follows from $(a_n,t_n)\in\underline{M}$. In the second case, if $a_1$ is offered in $\underline{M}$ with some transfer $t_1$, suppose that for some sequence $(\epsilon_n)_{n=1}^{\infty}$ of positive numbers converging to zero, there is some critical action $a_n\in(a_1,a_1+\epsilon_n)$. However, $a_n\rightarrow a_1$ while $(a_1,t_1)\in \underline{M}$ and the critical plan with action $a_n$ is not in $\underline{M}$, again causing a discontinuity contradiction; more precisely, since $a_n>a_1$, the critical plan offering $a_n$ is not in $\underline{M}$, and as every critical action in $[a_l,a_1]$ belongs to $A_{\succ}$, there is a critical action $\widetilde{a}_n\in(a_1,a_n]$ with $\widetilde{a}_n\notin A_{\succ}$; moreover, $a_1<a$ because $a_1$ is offered in $\underline{M}$, which contains no plan with action $a'\geq a$, so $\widetilde{a}_n<a<a^*$ for large $n$, which implies $r(\widetilde{a}_n)\notin R(\widetilde{a}_n;M)$ and thus $\widetilde{a}_n\in A_{\prec}$; however, $\widetilde{a}_n\rightarrow a_1$ and $a_1$ belongs to the open set $A_{\succ}$, which is disjoint from $A_{\prec}$. Hence, there is some $\epsilon>0$ such that no critical action lies in $(a_1,a_1+\epsilon)$; since the set of critical actions in $[a_1+\epsilon,a]$ is compact (Lemma \ref{lemma-critical}) and contains the critical action $a>a_1$, there must exist a lowest critical action that is strictly larger than $a_1$, denoted $a_2$, offered with some transfer $t_2$. In this case, $r(a_1)\succ_{AI}R(a_1;M)$ and $r(a_2)\prec_{AI}R(a_2;M)$, where the latter holds because $(a_2,t_2)\notin\underline{M}$, no critical action lies in $(a_1,a_2)$, and the critical action $a_2\leq a<a^*$ cannot have $r(a_2)\in R(a_2;M)$.

Next, I argue that under $r(a_1)$, $(a_2,t_2)$ is weakly preferred by the agent to $(a_1,t_1)$. To see this, I consider the set of all outsider actions under which $(a_1,t_1)$ is optimal, and let the one generating the highest agent incentives be $r_1$ (it exists because this set is nonempty and, by the closedness of $G$ in the proof of Lemma \ref{lemma-critical}, compact), which must satisfy $r_1\in R(a_1;M)$ by Lemma \ref{lemma-r in R}. Therefore, $r(a_1)\succ_{AI}r_1\in R(a_1;M)$. Suppose the claim above is not true: $(a_1,t_1)$ is strictly better than $(a_2,t_2)$ under $r(a_1)$. Hence, the former will be even better compared to the latter under $r_1$ because $a_1<a_2$ and $r_1\prec_{AI}r(a_1)$. Another conclusion one can reach is that under $r_1$, the agent must be indifferent between $(a_1,t_1)$ and some $(a_3,t_3)$ with $a_3>a_2$. To see this, take a sequence $(r'_n)_{n=1}^{\infty}$ with $r'_n\succ_{AI}r_1$ and $h(r'_n)\downarrow h(r_1)$, which exists because $r(a_1)\succ_{AI}r_1$ and $h$ is continuous on the interval $R$. By the maximality of $r_1$, $(a_1,t_1)$ is not optimal given any $r'_n$. Let $(x_n,\tau_n)$ be a plan that is optimal given $r'_n$. By Lemma \ref{lemma-critical}, $x_n\geq a_1$ because $(a_1,t_1)$ is optimal given $r_1\prec_{AI}r'_n$, and $x_n\neq a_1$ because $(a_1,t_1)$ is the unique critical plan offering $a_1$ and it is not optimal given $r'_n$; hence, $x_n$ is a critical action larger than $a_1$, so $x_n\geq a_2$. As $M$ and $R$ are compact, along a subsequence, $(x_n,\tau_n)$ converges to some $(a_3,t_3)\in M$ and $r'_n$ converges to some $\widetilde{r}$ with $h(\widetilde{r})=h(r_1)$, that is, $\widetilde{r}\sim_{AI}r_1$. By the closedness of $G$ in the proof of Lemma \ref{lemma-critical}, $(a_3,t_3)$ is optimal given $\widetilde{r}$, and thus also given $r_1$ by Lemma \ref{lemma-class}, so the agent is indifferent between $(a_1,t_1)$ and $(a_3,t_3)$ under $r_1$. Moreover, $a_3\geq a_2$, and $a_3\neq a_2$ because the critical plan $(a_3,t_3)$ is optimal given $r_1$, whereas $(a_2,t_2)$, the unique critical plan offering $a_2$, is strictly worse than $(a_1,t_1)$ under $r_1$. Therefore, $a_3>a_2$. A problem then emerges. That is, since $a_3>a_2>a_1$ and $a_2$ is suboptimal given $r_1$, under any $r'\succsim_{AI}r_1$, $(a_3,t_3)$ is strictly better than $(a_2,t_2)$ while under any $r'\precsim_{AI}r_1$, $(a_1,t_1)$ is strictly better than $(a_2,t_2)$. This means $(a_2,t_2)$ is never a best reply of the agent, which contradicts it being critical. At this point, I have shown $(a_2,t_2)$ is weakly preferred by the agent to $(a_1,t_1)$ under $r(a_1)$, and the symmetric arguments will show that $(a_1,t_1)$ is weakly preferred by the agent to $(a_2,t_2)$ under $r(a_2)$.

With these two facts, the continuity of $r(\cdot)$ with respect to the agent's mixed strategy then yields the existence of a strategy $\beta$ that mixes $(a_1,t_1)$ and $(a_2,t_2)$, and makes the agent indifferent between them under $r(\beta)$. I hence claim that $(\beta,r(\beta))$ is an equilibrium. It suffices to show that one cannot have a profitable deviation $(a',t')$ with $a'>a_2$, because the case with $a'<a_1$ is symmetric; also, the best replies to $r(\beta)$ must lie outside $(a_1,a_2)$ since, otherwise, there would be a critical plan in this interval, which contradicts $a_1$ and $a_2$ being adjacent critical actions. To see that such a deviation cannot happen, note that under any $r'\succsim_{AI}r(\beta)$, $(a_2,t_2)$ is still worse than $(a',t')$, while under any $r'\prec_{AI}r(\beta)$, $(a_2,t_2)$ becomes worse than $(a_1,t_1)$ due to their indifference under $r(\beta)$. Therefore, I obtain a contradiction to $(a_2,t_2)$ being a best reply to some $r$. At this point, I have reached a contradiction to the supposition of Step 1. \\

Step 2: suppose there is $a\in[a_0,a^*)$ with $\overline{r}(a)\in R(a;M)$ and $\overline{r}(a)\prec_{AI}r''$ for some $r''\in R(a;M)$.

I apply the proof in Step 1 to discuss three cases. First, if $a$ is critical and $r(a)\sim_{AI}\overline{r}(a)\in R(a;M)$, Lemma \ref{lemma-equilibrium} (together with Lemma \ref{lemma-class}, which gives $r(a)\in R(a;M)$) implies that $a$ forms an equilibrium. Second, if $a$ is critical and $r(a)\prec_{AI}\overline{r}(a)\in R(a;M)$, then either $r(a)\in R(a;M)$, in which case Lemma \ref{lemma-equilibrium} again gives an equilibrium at $a$, or $r(a)\prec_{AI}R(a;M)$ by Lemma \ref{lemma-convexity} and Lemma \ref{lemma-class}; in the latter case, I then find a critical plan $(a,t)\in M$ with $a<a^*$ and $r(a)\prec_{AI}R(a;M)$. This establishes an intermediate result in the proof in Step 1, a result that is sufficient for the existence of an under-action equilibrium. Lastly, suppose $a$ is not critical. In this case, since $\overline{r}(a)\prec_{AI}r''$, Lemma \ref{lemma-two sides} gives a critical plan with action $a''\leq a$ that is optimal given $r''$. One also has $a''<a$ because $a$ is not critical. Three facts are thus obtained: a critical plan that offers $a''$ has $a''<a$, the plan is optimal given $r''$, and $r''\in R(a;M)$. Note that the first paragraph of Step 1 already showed that these lead to an under-action equilibrium.

Finally, the arguments above establish the following lemma for any contract.
\begin{lemma}\label{lemma-directional}
    For all $M\in\mathcal{M}$ and every critical plan $(a',t')$ in $M$, if $r(a')\prec_{AI}r'$ for some $r'\in R(a';M)$, then $M$ induces an equilibrium in which every on-path action is no greater than $a'$; if $r(a')\succ_{AI}r'$ for some $r'\in R(a';M)$, then $M$ induces an equilibrium in which every on-path action is no lower than $a'$.
\end{lemma}

In fact, Step 1 above proves the first part of Lemma \ref{lemma-directional}. To see this, I fix $M\in\mathcal{M}$ and a critical plan $(a',t')$ with $r(a')\prec_{AI}r'$ for some $r'\in R(a';M)$. If $r(a')\in R(a';M)$, Lemma \ref{lemma-equilibrium} implies that $(a',t')$ forms an equilibrium, whose on-path action is $a'$. Otherwise, Lemma \ref{lemma-convexity} and Lemma \ref{lemma-class} yield $r(a')\prec_{AI}R(a';M)$, so $(a',t')$ is a critical plan with $r(a')\prec_{AI}R(a';M)$, as is the plan $(a,t)$ found in the first paragraph of Step 1. Then the arguments of Step 1, from the paragraph on the critical plan with the lowest action onward, apply with $(a',t')$ in the role of $(a,t)$. These arguments appeal to full implementation only to exclude a critical plan $(\widetilde{a},\widetilde{t})$ with $\widetilde{a}\leq a'$ and $r(\widetilde{a})\in R(\widetilde{a};M)$, and such a plan itself forms an equilibrium by Lemma \ref{lemma-equilibrium}, with on-path action $\widetilde{a}\leq a'$. Hence, either some critical plan of this kind exists and forms an equilibrium as desired, or no such plan exists and the arguments go through verbatim, constructing an equilibrium in which the agent mixes $(a_1,t_1)$ and $(a_2,t_2)$ with $a_1<a_2\leq a'$. In both cases, every on-path action is no greater than $a'$. Note also that these arguments use neither the null plan $(a_0,0)$ nor Assumption \ref{intuitive mixture}. The second part of Lemma \ref{lemma-directional} follows from the symmetric arguments, which reverse both the order of actions and the order $\succsim_{AI}$ and start from the critical plan with the highest action. By Lemma \ref{lemma-critical}, this plan is optimal given $\max_{AI}R$.

\subsection{Proof of Proposition \ref{proposition-upper bound}}

Fix any $M$ that fully implements $\alpha$. In the unique equilibrium, the outsider plays $r(\alpha)$, and every plan $(a,t)\in\supp(\beta_M)$ achieves $V(r(\alpha);M)$, so Lemma \ref{lemma-r in R} gives $T(a;M)=t$ and thus $V(r(\alpha);M)=u_A(a,r(\alpha))-T(a;M)$. Since every $a\in\supp(\alpha)$ is the action of such a plan (see the proof of Claim \ref{claim 3}), I obtain $T(a;M)-T(\underline{a}(\alpha);M)=u_A(a,r(\alpha))-u_A(\underline{a}(\alpha),r(\alpha))$ for all $a\in\supp(\alpha)$. Substituting this into Claim \ref{claim 1} gives
\[U_P(M)=\int_A\left[u_P(a,r(\alpha))+u_A(a,r(\alpha))\right]d\alpha(a)-u_A(\underline{a}(\alpha),r(\alpha))+T(\underline{a}(\alpha);M)=U_0(\alpha)+T(\underline{a}(\alpha);M).\]
By Claim \ref{claim 2}, $T(\underline{a}(\alpha);M)=T(a_0;M)+\int_{a_0}^{\underline{a}(\alpha)}\frac{\partial}{\partial a}u_A(a,\widetilde{r}(a))da$ with each $\widetilde{r}(a)\in R(a;M)$. For every $a<\underline{a}(\alpha)$, Claims \ref{claim 3} and \ref{claim 4} give $\widetilde{r}(a)\precsim_{AI}r(\alpha)$ and $\widetilde{r}(a)\precsim_{AI}\overline{r}(a)$, that is, $\widetilde{r}(a)\precsim_{AI}r^*_{\alpha}(a)$, so $\frac{\partial}{\partial a}u_A(a,\widetilde{r}(a))\leq\frac{\partial}{\partial a}u_A(a,r^*_{\alpha}(a))$ by the definition of $\succsim_{AI}$. Therefore, $U_P(M)\leq U_0(\alpha)+\overline{T}(\alpha)+T(a_0;M)$. At this point, to obtain the upper bound (\ref{upper bound}), it remains to show $T(a_0;M)\leq0$. By (\ref{T>t}) in the proof of Claim \ref{claim 1}, $T(a_0;M)\leq t$ for all $(a_0,t)\in M$, so $T(a_0;M)\leq0$ because $(a_0,0)\in M$. Finally, (\ref{upper bound}) results from combining this with the above inequality, since $M\in\mathcal{M}_U(\alpha)$ is arbitrary.

\subsection{Proof of Theorem \ref{main result}}

This proof is done by verifying that the contract stated in Theorem \ref{main result} approximately achieves the upper bound on the principal's payoff (\ref{upper bound}) in the unique equilibrium, for any fully implementable $\alpha\in\mathcal{A}_{FI}$. Again, for notational conciseness, I use $a^*$ to replace $\underline{a}(\alpha)$.

To start with, I construct an approximating sequence $(M_n)_{n=1}^{\infty}$. It then suffices to show that each $M_n$ fully implements $(\alpha,r(\alpha))$ and their payoff guarantee $U_P(M_n)$ converges to the upper bound (\ref{upper bound}), which implies $U_P^*(\alpha)=$ (\ref{upper bound}). Taking a small $\varepsilon>0$, I construct $M_n$ as
\[M_n:=\{(a,t^*_{\alpha}(a)-(a-a_0)\frac{\varepsilon}{n}):a\in A^*_{\alpha}\text{ and }a<a^*\}\cup\{(a,t^*_{\alpha}(a)-(a^*-a_0)\frac{\varepsilon}{n}):a\in A^*_{\alpha}\text{ and }a\geq a^*\}.\]
That is, compared to $M^*(\alpha)$, $M_n$ slightly lowers the marginal transfer at each off-path action by $\frac{\varepsilon}{n}$, while the transfers for the on-path actions are uniformly reduced by $(a^*-a_0)\frac{\varepsilon}{n}$. Each $M_n$ is a feasible contract in $\mathcal{M}$. First, $M_n$ contains the null plan $(a_0,0)$ because $a_0\in A^*_{\alpha}$ and $t^*_{\alpha}(a_0)=0$. Second, $M_n$ has a compact graph. To see this, note that the function $\phi(a):=h(r(a))-\min\{\max_{a'\in[a_0,a]}h(r(a')),h(r(\alpha))\}$ is continuous on $A$ and satisfies $\phi(a)=h(r(a))-h(r^*_{\alpha}(a))$ for all $a<a^*$, so the closure of $\{a<a^*:r(a)\sim_{AI}r^*_{\alpha}(a)\}$ is contained in $\{a\leq a^*:\phi(a)=0\}$ and thus adds at most $a^*\in\supp(\alpha)$ to this set. Since $\supp(\alpha)$ is also closed, $A^*_{\alpha}$ is closed, and $M_n$ is the graph of the continuous function $a\mapsto t^*_{\alpha}(a)-(\min\{a,a^*\}-a_0)\frac{\varepsilon}{n}$ on the compact set $A^*_{\alpha}$. The same argument with $\varepsilon=0$ shows that $M^*(\alpha)\in\mathcal{M}$. One can easily see that $M_n$ converges to $M^*(\alpha)$ with respect to the Hausdorff metric, since both contracts offer the actions in $A^*_{\alpha}$ and their transfers differ by at most $(a^*-a_0)\frac{\varepsilon}{n}$.

Next, I take any agent strategy $\beta$ and demonstrate that if the lowest action chosen in $\beta$ is below $a^*$, $(\beta,r(\beta))$ is not an equilibrium. To do this, let $A(\beta):=\{a':(a',t')\in\supp(\beta)\}$ and let the minimal action be $a_l:=\min A(\beta)<a^*$. I first show $r(\beta)\succsim_{AI}r^*_{\alpha}(a_l)$. By construction, every $a\in A^*_{\alpha}$ with $a<a^*$ satisfies $r(a)\succsim_{AI}r^*_{\alpha}(a)$: one has $r(a)\sim_{AI}r^*_{\alpha}(a)$ if $a\neq a_0$, and $r^*_{\alpha}(a_0)=\min_{AI}\{r(a_0),r(\alpha)\}\precsim_{AI}r(a_0)$. Moreover, $r^*_{\alpha}$ is nondecreasing in the order $\succsim_{AI}$, and $r^*_{\alpha}(a)\precsim_{AI}r(\alpha)$ for all $a$. For the on-path actions, the necessity part of Proposition \ref{proposition-implementability}, whose proof does not use Theorem \ref{main result}, gives $|\supp(\alpha)|\leq2$ and $r(a^*)\succsim_{AI}r(\alpha)\succsim_{AI}r(\overline{a}(\alpha))$. I discuss two cases. First, if $r(\beta)\succsim_{AI}r(\alpha)$, then $r(\beta)\succsim_{AI}r^*_{\alpha}(a_l)$ follows immediately. Second, suppose $r(\beta)\prec_{AI}r(\alpha)$. If $\overline{a}(\alpha)>a^*$, the transfers of the plans offering $\overline{a}(\alpha)$ and $a^*$ in $M_n$ differ by $t^*_{\alpha}(\overline{a}(\alpha))-t^*_{\alpha}(a^*)=u_A(\overline{a}(\alpha),r(\alpha))-u_A(a^*,r(\alpha))$, so under $r(\beta)$, the agent's payoff from the former plan minus that from the latter is $\int_{a^*}^{\overline{a}(\alpha)}\left[\frac{\partial}{\partial a}u_A(s,r(\beta))-\frac{\partial}{\partial a}u_A(s,r(\alpha))\right]ds<0$ by Assumption \ref{ranked incentives}, and thus $\overline{a}(\alpha)\notin A(\beta)$. Hence, every $a\in A(\beta)$ satisfies $r(a)\succsim_{AI}r^*_{\alpha}(a_l)$: if $a<a^*$, then $r(a)\succsim_{AI}r^*_{\alpha}(a)\succsim_{AI}r^*_{\alpha}(a_l)$ as $a\geq a_l$; if $a=a^*$, then $r(a)\succsim_{AI}r(\alpha)\succsim_{AI}r^*_{\alpha}(a_l)$. In the second case, it suffices to show that $r(\beta)$ lies between $\min_{a\in A(\beta)}r(a)$ and $\max_{a\in A(\beta)}r(a)$ in the order of $R$, both of which exist because $A(\beta)$ is compact and $r(\cdot)$ is continuous. Indeed, one must have $r(\beta)\geq\min_{a\in A(\beta)}r(a)$ because, otherwise, $r(\beta)<r(a)$ for all $a\in A(\beta)$, which yields the following
\[\frac{\partial}{\partial r}\int_{M_n}u_O(a,r(\beta))d\beta(a)=\int_{M_n}\frac{\partial}{\partial r}u_O(a,r(\beta))d\beta(a)>\int_{M_n}\frac{\partial}{\partial r}u_O(a,r(a))d\beta(a)\geq0,\]
where I apply the strict concavity of $u_O$ in $r$ and the fact that, for each $a\in A(\beta)$, $r(a)$ is not the lower corner of $R$ as $r(a)>r(\beta)$, so the optimality of $r(a)$ requires $\frac{\partial}{\partial r}u_O(a,r(a))\geq0$. This contradicts the optimality of $r(\beta)$, which is not the upper corner of $R$ because $r(\beta)<r(a)$ for all $a\in A(\beta)$. Symmetrically, one also has $r(\beta)\leq\max_{a\in A(\beta)}r(a)$. Finally, the quasi-concavity of $\succsim_{AI}$ (Assumption \ref{intuitive mixture}) indicates that since $r^*_{\alpha}(a_l)\precsim_{AI}r(a)$ for all $a\in A(\beta)$, one also has $r^*_{\alpha}(a_l)\precsim_{AI} r$ for all $r$ such that $r\geq\min_{a\in A(\beta)}r(a)$ and $r\leq\max_{a\in A(\beta)}r(a)$, which, along with the above, yields $r(\beta)\succsim_{AI}r^*_{\alpha}(a_l)$. Given outsider reply $r(\beta)$ and the assumption that $a_l<a^*$, I obtain a profitable deviation for the agent from $a_l$. For each $a\in A^*_{\alpha}$, let $U_n(a):=u_A(a,r(\beta))-t^*_{\alpha}(a)+(\min\{a,a^*\}-a_0)\frac{\varepsilon}{n}$ be the agent's payoff under $r(\beta)$ from the plan offering $a$ in $M_n$. For every $a\in A^*_{\alpha}$ with $a>a_l$, the definition of $t^*_{\alpha}$ in (\ref{optimal contract}) gives
\begin{equation*}
    U_n(a)-U_n(a_l)=\int_{a_l}^{a}\left[\frac{\partial}{\partial a}u_A(s,r(\beta))-\frac{\partial}{\partial a}u_A(s,r^*_{\alpha}(s))\right]ds+(\min\{a,a^*\}-a_l)\frac{\varepsilon}{n}.
\end{equation*}
I discuss two exhaustive cases. First, suppose that $a_l$ is the limit of a decreasing sequence of offered actions $a'_k\in A^*_{\alpha}\cap(a_l,a^*)$. Since $r(\beta)\succsim_{AI}r^*_{\alpha}(a_l)$, the integrand in the above display is at least $\frac{\partial}{\partial a}u_A(s,r^*_{\alpha}(a_l))-\frac{\partial}{\partial a}u_A(s,r^*_{\alpha}(s))$, which tends to $0$ as $s\downarrow a_l$. To see this, recall that $\frac{\partial}{\partial a}u_A(s,\cdot)$ represents $\succsim_{AI}$ for each fixed $s$ by Assumption \ref{ranked incentives}, so\[\frac{\partial}{\partial a}u_A(s,r^*_{\alpha}(s))=\min\left\{\max_{a'\in[a_0,s]}\frac{\partial}{\partial a}u_A(s,r(a')),\frac{\partial}{\partial a}u_A(s,r(\alpha))\right\}\]for all $s<a^*$, which is continuous in $s$ by Berge's maximum theorem. Hence, there is $\delta\in(0,a^*-a_l]$ such that the integrand exceeds $-\frac{\varepsilon}{2n}$ for all $s\in[a_l,a_l+\delta)$, and every $a'_k<a_l+\delta$ satisfies $U_n(a'_k)-U_n(a_l)\geq(a'_k-a_l)\frac{\varepsilon}{2n}>0$. Second, if the first case fails, then $A^*_{\alpha}\cap(a_l,a_l+\eta)=\emptyset$ for some $\eta>0$. Since $A^*_{\alpha}$ is closed and contains $a^*$, the set $A^*_{\alpha}\cap(a_l,a^*]=A^*_{\alpha}\cap[a_l+\eta,a^*]$ is compact and nonempty, so the next offered action $a^+:=\min(A^*_{\alpha}\cap(a_l,a^*])$ exists and satisfies $a^+>a_l$. I claim that $r^*_{\alpha}(s)\sim_{AI}r^*_{\alpha}(a_l)$ for all $s\in[a_l,a^+)$. Otherwise, since $r^*_{\alpha}$ is nondecreasing, some $s_1\in(a_l,a^+)$ has $h(r^*_{\alpha}(s_1))>h(r^*_{\alpha}(a_l))$, so $h(r^*_{\alpha}(a_l))<h(r(\alpha))$ and thus $h(r^*_{\alpha}(a_l))=h(\overline{r}(a_l))$. Take any $v$ with $h(\overline{r}(a_l))<v<h(r^*_{\alpha}(s_1))$. Since $h(r^*_{\alpha}(s_1))\leq h(\overline{r}(s_1))=\max\{h(\overline{r}(a_l)),\max_{a'\in[a_l,s_1]}h(r(a'))\}$, one has $h(r(a_l))<v<\max_{a'\in[a_l,s_1]}h(r(a'))$, so by the intermediate value theorem, there is a smallest $s_v\in[a_l,s_1]$ with $h(r(s_v))=v$. Then, $s_v>a_l$, and $h(r(a'))<v$ for all $a'\in[a_l,s_v)$ by the minimality of $s_v$, so $h(\overline{r}(s_v))=v<h(r(\alpha))$. As a result, $r(s_v)\sim_{AI}\overline{r}(s_v)\sim_{AI}r^*_{\alpha}(s_v)$ with $a_l<s_v<a^+\leq a^*$, that is, $s_v\in A^*_{\alpha}\cap(a_l,a^+)$, which contradicts the definition of $a^+$. Therefore, $r^*_{\alpha}(s)\sim_{AI}r^*_{\alpha}(a_l)\precsim_{AI}r(\beta)$ for all $s\in[a_l,a^+)$, so the integrand in the above display is nonnegative on $[a_l,a^+)$. Since $a^+\leq a^*$, the last term in the above display equals $(a^+-a_l)\frac{\varepsilon}{n}$ at $a=a^+$, so deviating from $a_l$ to $a^+$ increases the agent's payoff, that is,
\begin{equation*}
    U_n(a^+)-U_n(a_l)\geq(a^+-a_l)\frac{\varepsilon}{n}>0.
\end{equation*}
In both cases, the agent strictly prefers another plan in $M_n$ to the plan offering $a_l$ under $r(\beta)$, so $(\beta,r(\beta))$ is not an equilibrium. This wraps up the first part of the proof, which disproves the existence of any bad equilibrium that involves actions outside $\supp(\alpha)$.

In addition, I explain why, other than $(\alpha,r(\alpha))$, no equilibrium in which the agent chooses only actions in $\supp(\alpha)$ can exist. Suppose that such an equilibrium exists where the agent plays $\alpha'\neq\alpha$. Then, this will imply that $\alpha$ is not fully implementable. Since $\alpha'\neq\alpha$, $\alpha'\in\Delta(\supp(\alpha))$, and $|\supp(\alpha)|\leq2$ by Proposition \ref{proposition-implementability}, $\alpha$ is not degenerate, and there are three exhaustive cases to discuss. First, $\alpha'$ may assign positive probability to both $\underline{a}(\alpha)$ and $\overline{a}(\alpha)$, so the agent is indifferent between the two on-path plans given $r(\alpha')$. Since he is also indifferent between them given $r(\alpha)$, as I verify at the end of this proof, one has $\int_{\underline{a}(\alpha)}^{\overline{a}(\alpha)}\left[\frac{\partial}{\partial a}u_A(s,r(\alpha'))-\frac{\partial}{\partial a}u_A(s,r(\alpha))\right]ds=0$, and Assumption \ref{ranked incentives} yields $r(\alpha')\sim_{AI}r(\alpha)$. I note that whenever $\alpha$ forms an equilibrium under some contract, the agent is indifferent among the actions in $\supp(\alpha)$ given $r(\alpha)$. As a result, by Lemma \ref{lemma-class}, the on-path plans remain optimal given $r(\alpha')$, implying another equilibrium. Second, $\alpha'$ may be degenerate at the highest action $\overline{a}(\alpha)$, in which case one has $r(\overline{a}(\alpha))\succ_{AI}r(\alpha)$. Indeed, the agent weakly prefers the plan offering $\overline{a}(\alpha)$ to that offering $\underline{a}(\alpha)$ given $r(\overline{a}(\alpha))$, that is, $\int_{\underline{a}(\alpha)}^{\overline{a}(\alpha)}\left[\frac{\partial}{\partial a}u_A(s,r(\overline{a}(\alpha)))-\frac{\partial}{\partial a}u_A(s,r(\alpha))\right]ds\geq0$, so Assumption \ref{ranked incentives} gives $r(\overline{a}(\alpha))\succsim_{AI}r(\alpha)$, while the third condition of Proposition \ref{proposition-implementability} excludes $r(\overline{a}(\alpha))\sim_{AI}r(\alpha)$ as $\alpha'\neq\alpha$. Symmetrically, the third case is that $\alpha'$ is degenerate at $\underline{a}(\alpha)$ and $r(\underline{a}(\alpha))\prec_{AI}r(\alpha)$. To address these two situations, take any contract $M$ that fully implements $\alpha$. Its on-path plans offering $\underline{a}(\alpha)$ and $\overline{a}(\alpha)$ are critical as they are optimal given $r(\alpha)$, and Lemma \ref{lemma-r in R} gives $r(\alpha)\in R(\underline{a}(\alpha);M)$ and $r(\alpha)\in R(\overline{a}(\alpha);M)$. In the third case, $r(\underline{a}(\alpha))\prec_{AI}r(\alpha)$, so the first part of Lemma \ref{lemma-directional} yields an equilibrium of $M$ in which every on-path action is at most $\underline{a}(\alpha)$. In the second case, $r(\overline{a}(\alpha))\succ_{AI}r(\alpha)$, so the second part of Lemma \ref{lemma-directional} yields an equilibrium of $M$ in which every on-path action is at least $\overline{a}(\alpha)$. Since $\underline{a}(\alpha)<\overline{a}(\alpha)$, either equilibrium differs from $(\alpha,r(\alpha))$, which contradicts the full implementation of $\alpha$ by $M$.

To finish the proof, I next verify that $(\alpha,r(\alpha))$ forms an equilibrium. This results from $r^*_{\alpha}(s)\precsim_{AI}r(\alpha)$ for all $s$, which holds by construction of $r^*_{\alpha}$. On the one hand, for all $a\in\supp(\alpha)$ and $a'\in A^*_{\alpha}$ with $a'<a^*$, $a$ is better than $a'$ given $r(\alpha)$
\begin{equation*}
    \begin{aligned}
        u_A(a,r(\alpha))&-t^*_{\alpha}(a)+(a^*-a_0)\frac{\varepsilon}{n}-\left[u_A(a',r(\alpha))-t^*_{\alpha}(a')+(a'-a_0)\frac{\varepsilon}{n}\right] \\
        =&\int_{a'}^{a}\left[\frac{\partial}{\partial a}u_A(s,r(\alpha))-\frac{\partial}{\partial a}u_A(s,r^*_{\alpha}(s))\right]ds+(a^*-a')\frac{\varepsilon}{n} \\
        \geq&\;(a^*-a')\frac{\varepsilon}{n}>0,
    \end{aligned}
\end{equation*}
where the first inequality is a result of $r(\alpha)\succsim_{AI}r^*_{\alpha}(s)$ for all $s$ and $a\geq a^*>a'$. On the other hand, the agent is indifferent among on-path plans given $r(\alpha)$. That is, for all $a_1,a_2\in\supp(\alpha)$,
\begin{equation*}
    \begin{aligned}
        u_A(a_1,r(\alpha))&-t^*_{\alpha}(a_1)+(a^*-a_0)\frac{\varepsilon}{n}-\left[u_A(a_2,r(\alpha))-t^*_{\alpha}(a_2)+(a^*-a_0)\frac{\varepsilon}{n}\right] \\
        =&\left[u_A(a_1,r(\alpha))-u_A(a_2,r(\alpha))\right]-\left[t^*_{\alpha}(a_1)-t^*_{\alpha}(a_2)\right]=0,
    \end{aligned}
\end{equation*}
where I use the fact that, by construction, $t^*_{\alpha}{'}(a')=\frac{\partial}{\partial a}u_A(a',r(\alpha))$ when $a'\in[\underline{a}(\alpha),\overline{a}(\alpha)]$. In summary, $M_n$ fully implements $\alpha$. Moreover, the on-path plans of $M_n$ are $(a,t^*_{\alpha}(a)-(a^*-a_0)\frac{\varepsilon}{n})$ with $a\in\supp(\alpha)$, where $t^*_{\alpha}(a)=\overline{T}(\alpha)+u_A(a,r(\alpha))-u_A(a^*,r(\alpha))$ because $r^*_{\alpha}=r(\alpha)$ on $[a^*,\overline{a}(\alpha)]$. Hence, by (\ref{U_P(M)}) and the definition of $U_0(\alpha)$ in (\ref{upper bound}),
\[U_P(M_n)=\int_A\left[u_P(a,r(\alpha))+t^*_{\alpha}(a)\right]d\alpha(a)-(a^*-a_0)\frac{\varepsilon}{n}=U_0(\alpha)+\overline{T}(\alpha)-(a^*-a_0)\frac{\varepsilon}{n},\]
which converges to $U_0(\alpha)+\overline{T}(\alpha)$. Since $U_P(M_n)\leq U_P^*(\alpha)\leq U_0(\alpha)+\overline{T}(\alpha)$ by Proposition \ref{proposition-upper bound}, I obtain $U_P^*(\alpha)=U_0(\alpha)+\overline{T}(\alpha)$ and $U_P(M_n)\rightarrow U_P^*(\alpha)$. As $M_n$ also converges to $M^*(\alpha)$ with respect to the Hausdorff metric, $M^*(\alpha)$ is robustly optimal for inducing $\alpha$.

\subsection[Proof of Corollary \ref{corollary-criterion}]{Proof of Corollary \ref{corollary-criterion}}

Fix any $\alpha\in\mathcal{A}_{FI}$, as in Section \ref{Main Result}. The corollary requires this, since for $\alpha\notin\mathcal{A}_{FI}$, $\mathcal{M}_U(\alpha)$ is empty and no contract is robustly optimal for inducing $\alpha$, even if $r(\alpha)\precsim_{AI}r(a_0)$. The partial-implementation solution in Footnote \ref{partial implementation} is the contract $M^P(\alpha):=\{(a_0,0)\}\cup\{(a,t_P(a)):a\in\supp(\alpha)\}$ with $t_P(a):=u_A(a,r(\alpha))-u_A(a_0,r(\alpha))$, and its value for the principal is
\[U_P^0(\alpha):=\int_A\left[u_P(a,r(\alpha))+t_P(a)\right]d\alpha(a)=U_0(\alpha)+\int_{a_0}^{\underline{a}(\alpha)}\frac{\partial}{\partial a}u_A(a,r(\alpha))da.\]
Recall from Proposition \ref{proposition-implementability} that $\supp(\alpha)\subset\{\underline{a}(\alpha),\overline{a}(\alpha)\}$ and $r(\underline{a}(\alpha))\succsim_{AI}r(\alpha)$.

I first show the ``if'' part. Suppose that $r(\alpha)\precsim_{AI}r(a_0)$. Then, $\overline{r}(a)\succsim_{AI}r(a_0)\succsim_{AI}r(\alpha)$ for all $a\in A$, so $r^*_{\alpha}(a)\sim_{AI}r(\alpha)$ for all $a\leq\overline{a}(\alpha)$. Hence, $\overline{T}(\alpha)=\int_{a_0}^{\underline{a}(\alpha)}\frac{\partial}{\partial a}u_A(a,r(\alpha))da$, and Proposition \ref{proposition-upper bound} gives $U_P^*(\alpha)\leq U_P^0(\alpha)$. Taking any $\varepsilon>0$, I construct
\[M_n:=\{(a_0,0)\}\cup\left\{\left(a,t_P(a)-\frac{\varepsilon}{n}\right):a\in\supp(\alpha)\right\},\]
and I call the plans in the second set on-path plans. Each $M_n$ is a finite set that contains $(a_0,0)$, so $M_n\in\mathcal{M}$, and $M_n$ converges to $M^P(\alpha)$ with respect to the Hausdorff metric because the transfers of the two contracts differ by at most $\frac{\varepsilon}{n}$. I show that $M_n$ fully implements $\alpha$. Given any outsider action $r$, the on-path plan offering $a$ gives the agent $u_A(a,r)-u_A(a,r(\alpha))+u_A(a_0,r(\alpha))+\frac{\varepsilon}{n}$, while the null plan gives him $u_A(a_0,r)$. In particular, under $r(\alpha)$, every on-path plan gives $u_A(a_0,r(\alpha))+\frac{\varepsilon}{n}$, which is more than what the null plan gives, so $(\alpha,r(\alpha))$ is an equilibrium. Now, take any equilibrium $(\beta,r(\beta))$, and let $A(\beta)$ be defined as in the proof of Theorem \ref{main result}. I first show $r(\beta)\succsim_{AI}r(\alpha)$. Suppose instead that $r(\beta)\prec_{AI}r(\alpha)$. If $\overline{a}(\alpha)>\underline{a}(\alpha)$, then under $r(\beta)$, the agent's payoff from the on-path plan offering $\overline{a}(\alpha)$ minus that from the on-path plan offering $\underline{a}(\alpha)$ is $\int_{\underline{a}(\alpha)}^{\overline{a}(\alpha)}\left[\frac{\partial}{\partial a}u_A(s,r(\beta))-\frac{\partial}{\partial a}u_A(s,r(\alpha))\right]ds<0$ by Assumption \ref{ranked incentives}, so $\overline{a}(\alpha)\notin A(\beta)$. Hence, $A(\beta)\subset\{a_0,\underline{a}(\alpha)\}$, where $r(a_0)\succsim_{AI}r(\alpha)$ and $r(\underline{a}(\alpha))\succsim_{AI}r(\alpha)$. As shown in the proof of Theorem \ref{main result}, $r(\beta)$ lies between $\min_{a\in A(\beta)}r(a)$ and $\max_{a\in A(\beta)}r(a)$ in the order of $R$, so Assumption \ref{intuitive mixture} yields $r(\beta)\succsim_{AI}r(\alpha)$, a contradiction. Therefore, $r(\beta)\succsim_{AI}r(\alpha)$, and under $r(\beta)$, the agent's payoff from the on-path plan offering $\underline{a}(\alpha)$ minus that from the null plan is
\[\int_{a_0}^{\underline{a}(\alpha)}\left[\frac{\partial}{\partial a}u_A(s,r(\beta))-\frac{\partial}{\partial a}u_A(s,r(\alpha))\right]ds+\frac{\varepsilon}{n}\geq\frac{\varepsilon}{n}>0.\]
Thus, the null plan is not chosen in $\beta$, and the agent chooses only on-path plans, among which he is indifferent given $r(\alpha)$. The argument in the proof of Theorem \ref{main result} for equilibria in which the agent chooses only actions in $\supp(\alpha)$ then applies without change and shows that the agent plays $\alpha$ in $\beta$. Hence, $M_n$ fully implements $\alpha$, and
\[U_P(M_n)=\int_A\left[u_P(a,r(\alpha))+t_P(a)-\frac{\varepsilon}{n}\right]d\alpha(a)=U_P^0(\alpha)-\frac{\varepsilon}{n}.\]
Since $U_P(M_n)\leq U_P^*(\alpha)\leq U_P^0(\alpha)$, I obtain $U_P(M_n)\rightarrow U_P^*(\alpha)=U_P^0(\alpha)$. Therefore, $M^P(\alpha)$ is robustly optimal for inducing $\alpha$.

I next show the ``only if'' part. Suppose that $r(\alpha)\succ_{AI}r(a_0)$. Then, $\underline{a}(\alpha)>a_0$ because, otherwise, $r(a_0)=r(\underline{a}(\alpha))\succsim_{AI}r(\alpha)$. Since $h(\overline{r}(a))=\max_{a'\in[a_0,a]}h(r(a'))$ is continuous in $a$ and $h(\overline{r}(a_0))=h(r(a_0))<h(r(\alpha))$, there is $\eta\in(0,\underline{a}(\alpha)-a_0)$ such that $r^*_{\alpha}(a)=\overline{r}(a)\prec_{AI}r(\alpha)$ for all $a\in[a_0,a_0+\eta]$. By Assumption \ref{ranked incentives}, $\frac{\partial}{\partial a}u_A(a,r^*_{\alpha}(a))<\frac{\partial}{\partial a}u_A(a,r(\alpha))$ for all $a\in[a_0,a_0+\eta]$, while $\frac{\partial}{\partial a}u_A(a,r^*_{\alpha}(a))\leq\frac{\partial}{\partial a}u_A(a,r(\alpha))$ for all $a<\underline{a}(\alpha)$. Hence, $\overline{T}(\alpha)<\int_{a_0}^{\underline{a}(\alpha)}\frac{\partial}{\partial a}u_A(a,r(\alpha))da$, and Proposition \ref{proposition-upper bound} gives $U_P^*(\alpha)<U_P^0(\alpha)$. Now, suppose that some sequence $(M_n)_{n=1}^{\infty}\subset\mathcal{M}_U(\alpha)$ converges to $M^P(\alpha)$ with respect to the Hausdorff metric. For each $n$, let $(\underline{a}(\alpha),t_n)\in M_n$ be a plan chosen in the unique equilibrium induced by $M_n$. As shown in the proof of Proposition \ref{proposition-upper bound}, $U_P(M_n)=U_0(\alpha)+T(\underline{a}(\alpha);M_n)=U_0(\alpha)+t_n$. Moreover, the Hausdorff convergence gives plans $(a_n,t'_n)\in M^P(\alpha)$ with $|a_n-\underline{a}(\alpha)|\rightarrow0$ and $|t'_n-t_n|\rightarrow0$. Since $M^P(\alpha)$ is finite and $\underline{a}(\alpha)>a_0$, one has $(a_n,t'_n)=(\underline{a}(\alpha),t_P(\underline{a}(\alpha)))$ for all large $n$, so $t_n\rightarrow t_P(\underline{a}(\alpha))$ and $U_P(M_n)\rightarrow U_0(\alpha)+t_P(\underline{a}(\alpha))=U_P^0(\alpha)$. This contradicts $U_P(M_n)\leq U_P^*(\alpha)<U_P^0(\alpha)$ for all $n$. Therefore, no sequence of fully implementing contracts converges to $M^P(\alpha)$, so $M^P(\alpha)$ is not robustly optimal for inducing $\alpha$.

\section{}\label{Appendix B}

\subsection{Proof of Proposition \ref{proposition-implementability}}

I fix any $\alpha\in\Delta(A)$. To begin with, I show the sufficiency of the three stated conditions. Consider a contract that charges sufficiently low transfers: that is, for a very high $t_0$, let $M_0=\{(a,u_A(a,r(\alpha))-u_A(a_0,r(\alpha))-t_0):a\in\supp(\alpha)\}\cup\{(a_0,0)\}$. Under every $r\in R$, a plan in $M_0$ other than the null plan, which I call a non-null plan, gives the agent $u_A(a,r)-u_A(a,r(\alpha))+u_A(a_0,r(\alpha))+t_0$ if it offers $a$, which exceeds his payoff $u_A(a_0,r)$ from the null plan because $t_0$ is very high. Hence, the null plan is never chosen. Moreover, by Lemma \ref{lemma-class}, the agent's ranking of plans in $M_0$ depends on $r$ only through its $\sim_{AI}$ class, so under every $r\sim_{AI}r(\alpha)$, he is indifferent among all non-null plans, as he is under $r(\alpha)$. In contrast, for all $a_1<a_2$ in $\supp(\alpha)$, the non-null plan offering $a_2$ gives him $\int_{a_1}^{a_2}\left[\frac{\partial}{\partial a}u_A(a',r)-\frac{\partial}{\partial a}u_A(a',r(\alpha))\right]da'$ more than the non-null plan offering $a_1$, so under every $r\succ_{AI}r(\alpha)$ [$r\prec_{AI}r(\alpha)$], Assumption \ref{ranked incentives} implies that his unique optimal plan is the non-null plan offering $\overline{a}(\alpha)$ [$\underline{a}(\alpha)$]. As one can readily verify, $\alpha$ forms an equilibrium. To check if there is another equilibrium, in which the agent must play some $\alpha'\in\Delta(\supp(\alpha))$ with $\alpha'\neq\alpha$ since the null plan is never chosen and distinct non-null plans offer distinct actions, I discuss three cases. First, one may have $r(\alpha')\sim_{AI}r(\alpha)$. However, this is ruled out by the last condition, $\{\alpha'\in\Delta(\supp(\alpha)):r(\alpha')\sim_{AI}r(\alpha)\}=\{\alpha\}$. Second, if $r(\alpha')\succ_{AI}r(\alpha)$, then $\alpha'$ must be degenerate at the highest action $\overline{a}(\alpha)$, so one has $r(\overline{a}(\alpha))\succ_{AI}r(\alpha)$. Symmetrically, the third case is that $r(\alpha')\prec_{AI}r(\alpha)$, so $\alpha'$ is degenerate at $\underline{a}(\alpha)$ and $r(\underline{a}(\alpha))\prec_{AI}r(\alpha)$. These two cases are not possible due to the second condition, $r(\underline{a}(\alpha))\succsim_{AI}r(\alpha)\succsim_{AI}r(\overline{a}(\alpha))$. Therefore, $M_0$ fully implements $\alpha$.

On the other hand, I show that full implementability implies these conditions. I take any $\alpha\in\mathcal{A}_{FI}$ and any $M\in\mathcal{M}_U(\alpha)$. Every plan $(a',t')$ chosen on the path of the unique equilibrium attains $V(r(\alpha);M)$, so it is critical, and Lemma \ref{lemma-r in R} gives $r(\alpha)\in R(a';M)$ and $t'=T(a';M)$; moreover, as shown in the proof of Claim \ref{claim 3}, every action in $\supp(\alpha)$ is offered by such an on-path plan. I then apply Lemma \ref{lemma-directional} to show the first condition, $|\supp(\alpha)|\leq2$. Suppose that there is some $a'\in\supp(\alpha)$ with $\underline{a}(\alpha)<a'<\overline{a}(\alpha)$. Then, one cannot have $r(a')\prec_{AI}r(\alpha)$ or $r(a')\succ_{AI}r(\alpha)$. This is because $r(\alpha)\in R(a';M)$ by Lemma \ref{lemma-r in R}, and an equilibrium with on-path actions no greater (or no lower) than $a'$, which exists by Lemma \ref{lemma-directional} applied to the on-path plan offering $a'$, must be different from $\alpha$, causing multiplicity. As a result, one must have $r(a')\sim_{AI}r(\alpha)$, which, however, sustains $a'$ as an undesired equilibrium: by Lemma \ref{lemma-class}, $R(a';M)$ is a union of $\sim_{AI}$-classes, so $r(\alpha)\in R(a';M)$ implies $r(a')\in R(a';M)$, and Lemma \ref{lemma-equilibrium} then implies that the on-path plan offering $a'$ forms an equilibrium, which differs from $\alpha$ since $\underline{a}(\alpha)<\overline{a}(\alpha)$. Next, the second condition, $r(\underline{a}(\alpha))\succsim_{AI}r(\alpha)\succsim_{AI}r(\overline{a}(\alpha))$, results from Lemma \ref{lemma-directional} applied to the on-path plans offering $\underline{a}(\alpha)$ and $\overline{a}(\alpha)$. That is, $r(\underline{a}(\alpha))\prec_{AI}r(\alpha)$ will lead to an equilibrium with actions no greater than $\underline{a}(\alpha)$, whereas $r(\overline{a}(\alpha))\succ_{AI}r(\alpha)$ implies an equilibrium with actions no lower than $\overline{a}(\alpha)$. If either case happens, $\alpha$ is not degenerate, as otherwise $r(\underline{a}(\alpha))=r(\alpha)=r(\overline{a}(\alpha))$, so the resulting equilibrium differs from $\alpha$ and full implementation is impossible. Finally, the third condition is straightforward as one cannot have $\alpha'\neq\alpha$ with $r(\alpha')\sim_{AI}r(\alpha)$ and $\supp(\alpha')\subset\supp(\alpha)$, which makes $\alpha'$ an equilibrium whenever $\alpha$ forms one: the agent can play each $a\in\supp(\alpha)$ with the on-path plan $(a,T(a;M))$, which is optimal under $r(\alpha)$ and thus, by Lemma \ref{lemma-class}, also under $r(\alpha')$.

\subsection{Proof of Proposition \ref{proposition-incentive attenuation}}

I take any outcomes $a_F\in A_F$ and $a_P\in A_P$, whose optimality in their associated problems implies the following incentive compatibility conditions
\begin{equation}\label{IC}
    \begin{aligned}
        U_0(a_F)+\int_{a_0}^{a_F}\frac{\partial}{\partial a}u_A(a',r^*_{a_F}(a'))da'&\geq U_0(a_P)+\int_{a_0}^{a_P}\frac{\partial}{\partial a}u_A(a',r^*_{a_P}(a'))da'; \\
        U_0(a_P)+\int_{a_0}^{a_P}\frac{\partial}{\partial a}u_A(a',r(a_P))da'&\geq U_0(a_F)+\int_{a_0}^{a_F}\frac{\partial}{\partial a}u_A(a',r(a_F))da'.
    \end{aligned}
\end{equation}
Adding the two inequalities and canceling repeated terms yield the following
\begin{equation}\label{z>z}
    \begin{aligned}
        &\int_{a_0}^{a_F}z_F(a')da'\geq\int_{a_0}^{a_P}z_P(a')da'\text{, where} \\
        &z_F(a):=\min\left\{0,\frac{\partial}{\partial a}u_A(a,\overline{r}(a))-\frac{\partial}{\partial a}u_A(a,r(a_F))\right\}; \\
        &z_P(a):=\min\left\{0,\frac{\partial}{\partial a}u_A(a,\overline{r}(a))-\frac{\partial}{\partial a}u_A(a,r(a_P))\right\}.
    \end{aligned}
\end{equation}

For the sake of contradiction, suppose $r(a_F)\succ_{AI}r(a_P)$. The first goal is to show that $a_F<a_P$ must follow from this supposition. To see this, notice that $r(a_F)\succ_{AI}r(a_P)$ implies $z_F(a)\leq z_P(a)$ and when $z_F(a)\neq0$, the inequality here is strict. Since I have assumed $r(a_F)\succ_{AI}r(a_0)$, $z_F(a)\neq0$ must occur in a nonzero measure region due to the continuity of $z_F(\cdot)$. Therefore, I obtain the following
\[\int_{a_0}^{a_F}z_F(a')da'<\int_{a_0}^{a_F}z_P(a')da'=\int_{a_0}^{a_P}z_P(a')da'+\int_{a_P}^{a_F}z_P(a')da'.\]
Combining this with (\ref{z>z}) gives $\int_{a_P}^{a_F}z_P(a')da'>0$. Note that $z_P(\cdot)\leq0$ by its construction, so this is impossible if $a_F\geq a_P$. Thus, I conclude $a_F<a_P$.

\begin{figure}[h]
    \centering
    \includegraphics[width=0.5\linewidth]{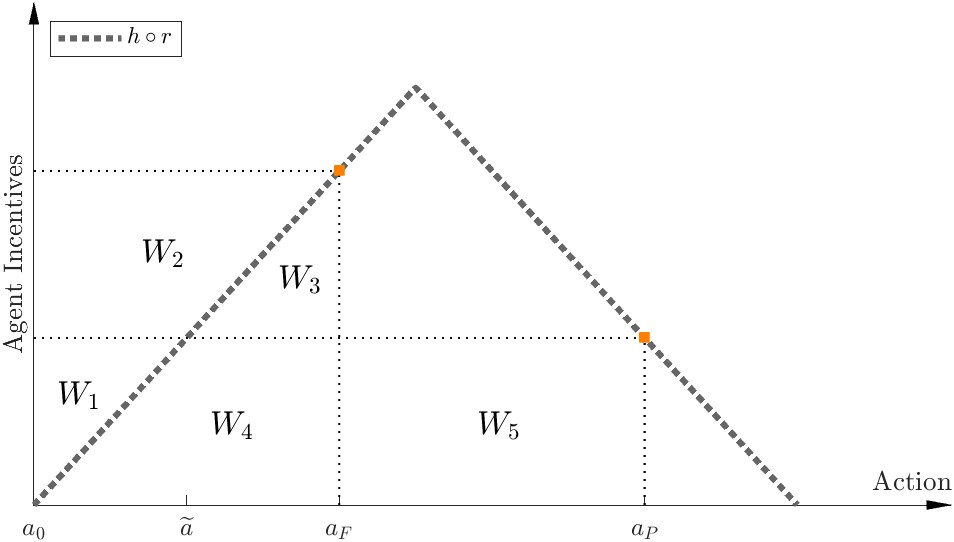}
    \caption{Decomposition in the proof of Proposition \ref{proposition-incentive attenuation}}
    \label{fig-decomp}
\end{figure}

Next, I show that incentive compatibility (\ref{IC}) cannot be maintained under both $r(a_F)\succ_{AI}r(a_P)$ and $a_F<a_P$. To do this, I find the lowest action $\widetilde{a}$ such that $r^*_{a_F}(\widetilde{a})\succsim_{AI}r(a_P)$. This is possible and $\widetilde{a}<a_F$ because $r(a_P)\prec_{AI}r(a_F)$ and $\succsim_{AI}$ is represented by a continuous function. In particular, if $r(a_P)\prec_{AI}r(a_0)$, $\widetilde{a}=a_0$, whereas, if $r(a_P)\succsim_{AI}r(a_0)$, $\widetilde{a}$ is the smallest action with $r^*_{a_F}(\widetilde{a})\sim_{AI}r(a_P)$. Now, I decompose the integrals in (\ref{IC}) into the following five parts
\begin{equation*}
    \begin{aligned}
        W_1&:=\int_{a_0}^{\widetilde{a}}\left[\frac{\partial}{\partial a}u_A(a',r(a_P))-\frac{\partial}{\partial a}u_A(a',r^*_{a_F}(a'))\right]da'; \\
        W_2&:=\int_{a_0}^{\widetilde{a}}\left[\frac{\partial}{\partial a}u_A(a',r(a_F))-\frac{\partial}{\partial a}u_A(a',r(a_P))\right]da'+\int_{\widetilde{a}}^{a_F}\left[\frac{\partial}{\partial a}u_A(a',r(a_F))-\frac{\partial}{\partial a}u_A(a',r^*_{a_F}(a'))\right]da'; \\
        W_3&:=\int_{\widetilde{a}}^{a_F}\left[\frac{\partial}{\partial a}u_A(a',r^*_{a_F}(a'))-\frac{\partial}{\partial a}u_A(a',r(a_P))\right]da'; \\
        W_4&:=\int_{a_0}^{\widetilde{a}}\frac{\partial}{\partial a}u_A(a',r^*_{a_F}(a'))da'+\int_{\widetilde{a}}^{a_F}\frac{\partial}{\partial a}u_A(a',r(a_P))da';\text{ and } W_5:=\int_{a_F}^{a_P}\frac{\partial}{\partial a}u_A(a',r(a_P))da'.
    \end{aligned}
\end{equation*}
To help the reader understand this decomposition, I illustrate these five parts in Figure \ref{fig-decomp}.

One caveat is that each area in Figure \ref{fig-decomp} only roughly represents, but may not equal, the corresponding value. This decomposition allows me to rewrite the principal's value, since $r^*_{a_P}(a')=r^*_{a_F}(a')$ for all $a'<\widetilde{a}$ and $r^*_{a_P}(a')=r(a_P)$ for all $a'\in[\widetilde{a},a_P]$. To see this, note that for $a'<\widetilde{a}$, one has $r^*_{a_F}(a')\prec_{AI}r(a_P)\prec_{AI}r(a_F)$, so $r^*_{a_F}(a')=\overline{r}(a')\prec_{AI}r(a_P)$ and thus $r^*_{a_P}(a')=\overline{r}(a')$, whereas for $a'\in[\widetilde{a},a_P)$, one has $\overline{r}(a')\succsim_{AI}\overline{r}(\widetilde{a})\succsim_{AI}r^*_{a_F}(\widetilde{a})\succsim_{AI}r(a_P)$, so $r^*_{a_P}(a')=r(a_P)$, which also holds at $a'=a_P$ by definition. For example, if she fully implements $a_F$, her value is $U_0(a_F)+W_3+W_4$ while the partial solution gives her $U_0(a_F)+W_1+W_2+W_3+W_4$. In this way, (\ref{IC}) becomes $U_0(a_F)+W_3+W_4\geq U_0(a_P)+W_4+W_5$ and $U_0(a_P)+W_1+W_4+W_5\geq U_0(a_F)+W_1+W_2+W_3+W_4$. Now, adding the two inequalities yields $W_2\leq0$. However, this is impossible because one must have $W_2>0$. The reason is that (i) $r(a_F)\succ_{AI}r(a_P)$ and $\widetilde{a}\geq a_0$ ensure that the first part of $W_2$'s definition above is not negative; (ii) one has $a_F>\widetilde{a}$, and for $a\in[\widetilde{a},a_F]$, $r(a_F)\succsim_{AI}r_{a_F}^*(a)$, and, if $\widetilde{a}>a_0$, in a nonzero measure region near $\widetilde{a}$, $r_{a_F}^*(\cdot)$ must be close to $r(a_P)$ due to its continuity, so $r(a_P)\prec_{AI}r(a_F)$ guarantees that the second part of $W_2$ is strictly positive; if $\widetilde{a}=a_0$, the first part of $W_2$ is zero, while the second part is still strictly positive because $r^*_{a_F}(a_0)\precsim_{AI}r(a_0)\prec_{AI}r(a_F)$ by the maintained assumption, so, by continuity, $r^*_{a_F}(\cdot)\prec_{AI}r(a_F)$ in a nonzero measure region near $a_0$.

As a result, $r(a_F)\succ_{AI}r(a_P)$ leads to a contradiction, so one must have $r(a_F)\precsim_{AI}r(a_P)$.

\end{document}